\documentclass[11pt, a4paper]{article}

\pdfoutput=1
\usepackage{jcappub}
\usepackage[T1]{fontenc}

\usepackage{appendix}
\usepackage{array}
\newcolumntype{x}[1]{>{\centering\arraybackslash}p{#1}}
\usepackage{bm}
\usepackage{color}
\usepackage{graphicx}
\usepackage{cancel}
\usepackage{amsfonts}
\usepackage{amssymb}
\usepackage{amsmath}
\usepackage{calc}
\usepackage{dcolumn}
\usepackage{mathrsfs}
\usepackage{mathtools}
\usepackage{soul}

\usepackage[normalem]{ulem} 

\newcommand{\eg}{e.g.~}
\newcommand{\ie}{i.e.~}

\newcommand{\Eq}[1]{Eq.~\eqref{#1}}
\newcommand{\Fig}[1]{Fig.~\ref{#1}}
\newcommand{\Sec}[1]{Sec.~\ref{#1}}

\newcommand{\Hcur}{\mathcal{H}}

\newcommand{\TeV}{\,{\rm TeV}}
\newcommand{\GeV}{\,{\rm GeV}}

\newcommand{\beq}{\begin{equation}}
\newcommand{\eeq}{\end{equation}}

\newcommand{\ud}{\text{d}}
\newcommand{\bol}[1]{\boldsymbol{#1}}

\newcommand{\ER}{E_\text{R}}
\newcommand{\Ed}{E'}

\newcommand{\vmin}{v_\text{min}}

\newcommand{\teta}{\tilde{\eta}}

\newcommand{\ignore}[1]{}

\definecolor{rossoCP3}{cmyk}{0,.88,.77,.40}
\definecolor{verdeCP3}{rgb}{0.09765625, 0.57421875, 0.1015625}
\definecolor{bluCP3}{rgb}{0, 0.23, 0.67}

\ifx\pdfoutput\undefined
\usepackage[dvips,bookmarks]{hyperref}    
\else
\usepackage{hyperref}    
\fi
\hypersetup{colorlinks, bookmarksopen, bookmarksnumbered,
citecolor=verdeCP3, linkcolor=bluCP3, pdfstartview=FitH, urlcolor=rossoCP3}

\notoc

\newcommand{\AddrUCLA}{Department of Physics and Astronomy, UCLA, 475 Portola Plaza, Los Angeles, CA 90095 (USA)}

\begin{document}

\title{\boldmath Updated Constraints on the Dark Matter Interpretation of CDMS-II-Si Data }

\author[a]{Samuel J.~Witte}
\author[a]{and Graciela B.~Gelmini}

\affiliation[a]{\AddrUCLA}

\emailAdd{switte@physics.ucla.edu}
\emailAdd{gelmini@physics.ucla.edu}

\abstract{We present an updated halo-dependent and halo-independent analysis of viable light WIMP dark matter candidates which could account for the excess observed in CDMS-II-Si. We include recent constraints from LUX, PandaX-II, and PICO-60, as well as projected sensitivities for XENON1T, SuperCDMS SNOLAB, LZ, DARWIN, DarkSide-20k, and PICO-250, on candidates with spin-independent isospin conserving and isospin-violating interactions, and either elastic or exothermic scattering. We show that there exist dark matter candidates which can explain the CDMS-II-Si data and remain very marginally consistent with the null results of all current experiments, however such models are highly tuned, making a dark matter interpretation of CDMS-II-Si very unlikely.  We find that these models can only be ruled out in the future by an experiment comparable to LZ or PICO-250. }

\keywords{dark matter theory, dark matter experiment}

\maketitle

\flushbottom

\newpage


\section{Introduction}

Despite an overwhelming amount of evidence for the existence of dark matter, very little is known about it beyond what is inferred from its gravitational influence. Motivated largely by theoretical expectations, weakly interacting massive particles (WIMPs) with mass at the $\GeV$ to $\simeq 100 \TeV$-scale remain among the most studied candidates.

Direct dark matter experiments search for the energy deposited into nuclei in underground detectors by collisions with WIMPs gravitationally bound to the galactic halo. While no definitive detections have been made, a number of collaborations have observed potential dark matter signals~\cite{Bernabei:2010mq,Aalseth:2010vx,Aalseth:2011wp,Aalseth:2012if,Aalseth:2014eft, Aalseth:2014jpa,Agnese:2013rvf}; however, such observations are typically viewed to be in conflict with the null results of many other experiments~\cite{Angle:2011th, Aprile:2011hi, Aprile:2012nq, Felizardo:2011uw, Archambault:2012pm, Behnke:2012ys, Ahmed:2012vq,Amole:2015lsj,Agnese:2015ywx,Akerib:2015rjg,Agnese:2015nto,Agnese:2014aze,Aprile:2016swn,Akerib:2016vxi,Tan:2016zwf,Amole:2017dex}.

The difficulty in making definitive statements regarding the nature of potentially viable signals arises from the fact that there exists a vast amount of uncertainty in the analysis of direct dark matter detection data. This is because both the particle physics and the astrophysics entering the computation of the expected scattering rates are, at best, poorly understood. In standard analyses of direct detection data, assumptions must be made on the local dark matter density, the dark matter velocity distribution, the dark matter-nuclei interaction, and the scattering kinematics. Bounds are then placed as a function of the dark matter mass and overall scale of the cross section. The obvious problem is that adjusting assumptions, \eg on the velocity distribution, unevenly alters the predicted rates in different experiments. This happens to be particularly true for the region of parameter space where potential dark matter signals have arisen, as this region sits near the low-energy threshold of many experiments. 

In recent years, `halo-independent' data comparison methods that avoid making any assumptions about the local dark matter halo characteristics have been developed, thereby reducing the uncertainty in experimental comparisons (see \eg\cite{Fox:2010bz,Fox:2010bu,Frandsen:2011gi,Gondolo:2012rs,HerreroGarcia:2012fu,Frandsen:2013cna,DelNobile:2013cta,Bozorgnia:2013hsa,DelNobile:2013cva,DelNobile:2013gba,DelNobile:2014eta,Feldstein:2014gza,Fox:2014kua,Gelmini:2014psa,Cherry:2014wia,DelNobile:2014sja,Scopel:2014kba,Feldstein:2014ufa,Bozorgnia:2014gsa,Blennow:2015oea,DelNobile:2015lxa,Anderson:2015xaa,Blennow:2015gta,Scopel:2015baa,Ferrer:2015bta,Wild:2016myz,Kahlhoefer:2016eds,Gelmini:2016pei,Kavanagh:2016xfi}). The original halo-independent analyses were rather limited in that putative signals often required averaging the signal over some energy range, potentially removing valuable information and making the comparison with upper limits ambiguous (see \eg \cite{Fox:2010bz,Frandsen:2011gi,Gondolo:2012rs,DelNobile:2013cva}). Recently, methods were developed which, for putative signals, allow for the construction of halo-independent confidence bands, resulting in a better comparison between upper limits and potential signals~\cite{Fox:2014kua,Gelmini:2015voa,Gelmini:2016pei}. These methods, however, rely on the ability to use an extended likelihood~\cite{Barlow:1990vc} for at least one of the experiments observing a putative signal. At the moment, CDMS-II-Si is the only experiment that has claimed a potential dark matter signal for which such a method can be applied.

Halo-dependent analyses strongly constrain the excess observed by CDMS-II-Si (see \eg \cite{DelNobile:2013gba,DelNobile:2014sja,Geng:2016uqt}). A halo-independent analysis performed on the CDMS-II-Si data in 2014 showed that the only WIMP candidates still consistent with the upper limits of null searches were those with spin-independent isospin-violating interactions, and either elastic or exothermic scattering~\cite{Gelmini:2015voa}. Here, we revisit the viability of the CDMS-II-Si excess, using both halo-dependent, assuming the standard halo model (SHM), and halo-independent analyses, incorporating the latest bounds produced by LUX (using their complete exposure)~\cite{Akerib:2016vxi}, PandaX-II~\cite{Tan:2016zwf}, and PICO-60~\cite{Amole:2017dex}. We also assess the projected sensitivity of XENON1T~\cite{Coderre:2014axc,Diglio:2016stt}, LZ~\cite{Akerib:2015cja,Szydagis:2016few}, DARWIN~\cite{Schumann:2015cpa}, DarkSide-20k~\cite{Agnes:2015ftt,Davini:2016vpd}, PICO-250~\cite{Pullia:2014vra}, and the high-voltage germanium detectors of SuperCDMS to be installed at SNOLAB~\cite{Agnese:2016cpb}. We show that models with highly exothermic kinematics and a neutron-to-proton coupling ratio $f_n/f_p$ set to minimize the scattering rate in xenon-target experiments are not currently excluded, nor can they be rejected by XENON1T.

In \Sec{sec:halorev} we review the halo-independent analysis and the procedure for constructing the two-sided pointwise halo-independent confidence band. The analysis for each experiment is explained in \Sec{sec:data}. In \Sec{sec:results}, we present our results, specifically focusing on isospin conserving and isospin-violating~\cite{Kurylov:2003ra,Feng:2011vu} (with $f_n/f_p=-0.8$ and $f_n/f_p=-0.7$) interactions with elastic and exothermic scattering~\cite{Finkbeiner:2007kk,Batell:2009vb,Graham:2010ca}. We conclude in \Sec{sec:conclu}.


\section{Halo-Independent Analysis \label{sec:halorev}}

\subsection{Halo-Independent Bounds}
Here, we briefly review the generalized halo-independent analysis implemented in \Sec{sec:results}, concentrating on the extended halo independent (EHI) analysis~\cite{Gelmini:2015voa} in the following subsection (the reader is encouraged to consult \cite{Gondolo:2012rs,DelNobile:2013cta,DelNobile:2013cva,DelNobile:2014sja,Gelmini:2014boa} for additional details). 

In direct detection experiments, the differential rate per unit detector mass of a target $T$, induced by collisions with a WIMP of mass $m$, as a function of nuclear recoil energy $\ER$ is given by
\begin{equation}
\label{diffrate}
\frac{\ud R_T}{\ud \ER} = \frac{\rho}{m}\frac{C_T}{m_T}\int_{v \geqslant \vmin(\ER)} \, \ud^3 \, v \, f(\bol{v},t) \, v \, \frac{\ud \sigma_T}{\ud \ER}(\ER, \bol{v}) \, ,
\end{equation}
where $m_T$ is the mass of the target element, $\rho$ is the local dark matter density, $C_T$ is the mass fraction of a nuclide $T$ in the detector, $\ud \sigma_T/\ud \ER$ is the dark matter-nuclide differential cross section in the lab frame, and $f(\bol{v},t)$ is the dark matter velocity distribution in the lab frame. The temporal dependence of $f(\bol{v},t)$ arises from Earth's rotation about the Sun. For the halo-dependent analyses in \Sec{sec:results}, we assume the SHM, \ie $f(\bol{v},t)$ is an isotropic Maxwellian velocity distribution in the Galactic frame, with the astrophysical parameters adopted in~\cite{Gelmini:2014psa}. 

The integration in \Eq{diffrate} runs over all dark matter particle speeds larger than or equal to $\vmin(\ER)$, the minimum speed necessary to impart an energy $\ER$ to the nucleus. Should multiple target nuclides be present in the detector, the total differential scattering rate is given by
\begin{equation}\label{sum_diffrate}
\frac{\ud R}{\ud \ER} = \sum_T \frac{\ud R_T}{\ud \ER} \, .
\end{equation}

For elastic scattering, the value of $\vmin$ is given by 
\begin{equation}
\vmin = \sqrt{\frac{m_T \ER}{2 \mu_T^2}} \, ,
\end{equation}
where $\mu_T$ is the WIMP-nuclide reduced mass. It may be possible that the dominant WIMP-nuclei interaction proceeds instead through an inelastic collision, whereby the dark matter particle $\chi$ scatters into a new state $\chi'$ with mass $m' = m + \delta$ (with $|\delta| \ll m$)~\cite{Finkbeiner:2007kk,Batell:2009vb,Graham:2010ca}. In the limit that $\mu_T |\delta|/m^2 \ll 1$, $\vmin(\ER)$ is instead given by
\begin{equation}\label{eq:vmin}
\vmin(\ER) = \frac{1}{\sqrt{2 m_T \ER}} \left| \frac{m_T \ER}{\mu_T} +\delta \right| \, ,
\end{equation}
where $\delta < 0$ ($\delta > 0$) corresponds to an exothermic (endothermic) scattering process. \Eq{eq:vmin} can be inverted to find the possible range of recoil energies which can be imparted by a dark matter particle with speed $v$ in the lab frame $\ER^{T,-} \leq \ER \leq \ER^{T,+}$, where 
\begin{equation}\label{eq:Ebranch}
\ER^{T,\pm} (v) = \frac{\mu_T^2 v^2}{2 m_T} \left( 1 \pm \sqrt{1-\frac{2\delta}{\mu_T v^2}} \right)^2 \, .
\end{equation}
It should be clear from \Eq{eq:Ebranch} that for endothermic scattering, for which $\delta > 0$, there exists a non-trivial kinematic endpoint for the WIMP speed given by $v_\delta^T = \sqrt{2 \delta / \mu_T} > 0$, such that dark matter particles traveling at speeds $v < v_\delta^T$ cannot induce nuclear recoils. In this paper we will be focusing exclusively on elastic ($\delta = 0$) and exothermic ($\delta < 0$) scattering, for which $v_\delta^T = 0$. Interpreting the CDMS-II-Si data using models with endothermic spin-independent interactions are clearly experimentally rejected. Notice that \Eq{eq:Ebranch} implies only a finite range of recoil energies around the energy $\ER(v_\delta^T) = \mu_T |\delta|/m_T$ can be probed for inelastic scattering.

Experiments do not directly measure the recoil energy of the nucleus, but rather a proxy for it that we denote $\Ed$. The differential rate in this new observable energy $\Ed$ is given by 
\begin{equation}\label{eq:diffrate_ep}
\frac{\ud R}{\ud \Ed} = \sum_T \int_0^\infty \ud \ER \, \epsilon(\ER,\Ed) \, G_T(\ER,\Ed) \, \frac{\ud R_T}{\ud \ER} \, ,
\end{equation}
where $\epsilon(\ER,\Ed)$ is the detection efficiency and $G_T(\ER,\Ed)$ is the energy resolution; jointly, these two functions give the probability that a detected recoil energy $\Ed$ resulted from a true nuclear recoil energy $\ER$.  

Changing the order of integration in \Eq{eq:diffrate_ep} allows the differential rate to be expressed as
\begin{equation}\label{diffrate_manip1}
\frac{\ud R}{\ud \Ed} = \frac{\sigma_\text{ref} \rho}{m} \int_{v \geqslant v_\delta^T} \ud^3 v \, \frac{f(\bol{v},t)}{v} \, \frac{\ud \Hcur}{\ud \Ed} (\Ed, \bol{v}) \, ,
\end{equation}
where we have defined
\begin{equation}\label{eq:dHcurl}
 \frac{\ud \Hcur}{\ud \Ed}(\Ed, \bol{v}) \equiv \sum_T
  \begin{dcases} 
      \hfill \frac{C_T}{m_T}\int_{\ER^{T,-}}^{\ER^{T,+}} \, \ud \ER \epsilon(\ER,\Ed) \, G_T(\ER, \Ed) \, \frac{v^2}{\sigma_\text{ref}} \, \frac{\ud \sigma_T}{\ud \ER}(\ER, \bol{v})    \hfill & \text{ if $v \geqslant v_\delta^T$,} \\
      \hfill 0 \hfill & \text{ if $v < v_\delta^T$\, .} \\
  \end{dcases}
\end{equation}
Here, we have explicitly factored out an overall normalization $\sigma_\text{ref}$ from the differential cross section. For spin-independent interactions, the differential WIMP-nucleus cross section is given by
\begin{equation}
\frac{\ud \sigma_T^{SI}}{\ud \ER}(\ER,v) = \sigma_p \frac{\mu_T^2}{\mu_p^2}[Z_T+(A_T-Z_T)(f_n/f_p)]^2 \, \frac{F_T^2(\ER)}{2 \mu_T^2 v^2 / m_T} \, ,
\end{equation}  
where $F_T(\ER)$ is the nuclear form factor that accounts for the decoherence of the dark matter-nuclide interaction at large momentum transfer. Here, we take this to be the Helm form factor~\cite{Helm:1956zz}. Thus we take $\sigma_\text{ref} = \sigma_p$, the WIMP-proton cross section. Interactions with spin- or nuclear magnetic moment-dependencies produce smaller rates in silicon relative to other target elements employed by experiments which have not observed an excess.

Let us define the halo function
\begin{equation}\label{eq:eta_t}
\tilde{\eta}(\vmin,t) \equiv \frac{\rho \sigma_\text{ref}}{m}\int_{\vmin}^{\infty} \, \ud v \, \frac{F(v,t)}{v} \, ,
\end{equation}   
where the function $F(v,t)$ is the local dark halo speed distribution, given by $F(v,t) = v^2 \int \ud \Omega_v f(\bol{v},t)$. Using \Eq{eq:eta_t}, the differential rate in $\Ed$ can be written as
\begin{equation}\label{drate_detatilde}
\frac{\ud R}{\ud \Ed} = - \int_{v_\delta}^{\infty} \ud v \, \frac{\partial \tilde{\eta}(v,t)}{\partial v} \, \frac{\ud \Hcur}{\ud \Ed}(\Ed, v) \, .
\end{equation}
Applying integration by parts on \Eq{drate_detatilde}, and noting that $\tilde{\eta}(\infty,t) = 0$ and $\ud \Hcur / \ud \Ed (\Ed, v_\delta) = 0$, the differential rate can be expressed as
\begin{equation}\label{eq:difres}
\frac{\ud R}{\ud \Ed} = \int_{v_\delta}^{\infty} \ud \vmin \tilde{\eta}(\vmin, t) \, \frac{\ud \mathcal{R}}{\ud \Ed}(\Ed, \vmin) \, ,
\end{equation}
where we have defined a WIMP model and experiment dependent ``differential response function'' $\ud \mathcal{R}/\ud \Ed$ as
\begin{equation}
\frac{\ud \mathcal{R}}{\ud \Ed}(\Ed,\vmin) \equiv \frac{\partial}{\partial \vmin}\left[ \frac{\ud \Hcur}{\ud \Ed}(\Ed,\vmin) \right] \, .
\end{equation} 

Approximating the time dependence of the halo function as
\begin{equation} \label{eta}
\tilde{\eta}(\vmin,t) \simeq \tilde{\eta}^0(\vmin) + \tilde{\eta}^1(\vmin) \cos(2 \pi (t-t_0)/\text{year}) \, ,
\end{equation}
and integrating \Eq{eq:difres} over $\Ed$, the unmodulated $R^0$ and annual modulation amplitude $R^1$ of the rate, integrated over an observable energy bin $[\Ed_1,\Ed_2]$, is given by 
\begin{align}
R^{\alpha}_{[\Ed_1,\Ed_2]} & \equiv \int_{v_\delta}^{\infty} \ud \vmin \, \tilde{\eta}^\alpha (\vmin) \, \int_{\Ed_1}^{\Ed_2} \, \ud \Ed \, \frac{\ud \mathcal{R}}{\ud \Ed}  \\ & \label{eq:rateeq}
 = \int_{v_\delta}^{\infty} \ud \vmin \, \tilde{\eta}^\alpha (\vmin) \, \mathcal{R}_{[\Ed_1,\Ed_2]}(\vmin)  \, ,
\end{align}
where $\alpha = 0,1$ and the second line has defined the energy integrated ``response function'' $\mathcal{R}$.

In order to place an upper limit on the function $\teta^0(\vmin)$ (hereby denoted $\teta(\vmin)$), we note that at a particular point in the $\vmin-\teta$ plane, the halo function producing the smallest number of events in a particular experiment is a downward step-function with the step located at the particular $(\vmin,\teta)$ point. This is a consequence of the fact that, by definition, $\teta(\vmin)$ is a monotonically decreasing function of $\vmin$. As first shown in~\cite{Fox:2010bz}, $90\%$ CL limits on $\teta$, $\teta^\text{lim}$, are placed by determining the $90\%$ CL limit on the rate, $R^\text{lim}_{[\Ed_1,\Ed_2]}$, and inverting \Eq{eq:rateeq}, \ie
\begin{equation}
\teta^\text{lim} (\vmin) = \frac{R^\text{lim}_{[\Ed_1,\Ed_2]}}{\int_{v_\delta}^{\vmin} \ud v \, \mathcal{R}_{[\Ed_1,\Ed_2]}(v)} \, .
\end{equation}

\subsection{Halo-Independent Confidence Band}

It was shown in \cite{Fox:2014kua,Gelmini:2015voa} that an extended likelihood is maximized by a piece-wise constant halo function $\teta_{BF}(\vmin)$ with a number of steps less than or equal to the number of events observed, and furthermore that a two-sided pointwise halo-independent confidence band can be constructed around this best-fit halo function, $\teta_{BF}$. A stream of velocity $\vec{v}_s$ with respect to the Galaxy, such that $|\vec{v}_s+\vec{v}_\oplus|=\vmin$ (where $\vec{v}_\oplus$ is Earth's velocity with respect to the Galaxy) would produce an $\teta$ function proportional to $\Theta(|\vec{v}_s+\vec{v}_\oplus|-\vmin)$. Thus a piecewise $\teta(\vmin)$ function could be interpreted as corresponding to a series of streams, one for each of its downward steps. More recently, it was shown that this formalism can be extended to more generalized likelihood functions that include at least one extended likelihood~\cite{Gelmini:2016pei}. Here, we briefly summarize the process outlined in \cite{Gelmini:2015voa} for producing a two-sided pointwise halo-independent confidence band using an extended likelihood function (which we apply in \Sec{sec:results} to the CDMS-II-Si data) of the form
\begin{equation}
\mathcal{L} = e^{-N_E[\teta]} \prod_{a=1}^{N_{\rm obs}}MT \frac{\ud R_{\rm tot}}{\ud \Ed}\Biggr|_{\Ed=E_a} \, ,
\end{equation}
where $N_E[\teta]$ is the total number of expected events, $N_{\rm obs}$ is the number of observed events, $\ud R_{\rm tot}/\ud \Ed$ is the total differential rate, and $\Ed_a$ is the detected energy of event $a$.

The confidence band is defined as the region in the $\vmin-\teta$ plane satisfying
\begin{equation}\label{CBdef}
\Delta L[\teta] \equiv L[\teta] - L_\text{min} \leq \Delta L^* \, ,
\end{equation}
where $L[\teta]$ is two times the minus log likelihood, $L_\text{min}$ is the value of $L[\teta]$ evaluated with the best-fit halo function $\teta_{BF}(\vmin)$, and $\Delta L^*$ corresponds to the desired confidence level. That is to say, we seek the collection of all halo functions that produce changes in the log likelihood function less than or equal to the desired value $\Delta L^*$. 

While this is a viable definition, in practice finding this complete set of halo functions is not possible. Instead, we consider the subset of $\teta$ functions which minimize $L[\teta]$ subject to the constraint 
\begin{equation}\label{constrain}
\teta(v^*) = \teta^* \, .
\end{equation}
We define $L_\text{min}^{c}(v^*,\teta^*)$ to be the minimum of $L[\teta]$ subject to the constraint in \Eq{constrain}, and we define the function $\Delta L_\text{min}^{c}(v^*,\teta^*)$ as 
\begin{equation}
\Delta L_\text{min}^{c}(v^*,\teta^*) \equiv L_\text{min}^{c}(v^*,\teta^*) - L_\text{min} \, .
\end{equation}
Should the point $(v^*,\teta^*)$ lie within the confidence band, then at least one halo function passing through this point should satisfy $\Delta L[\teta] \leq \Delta L^*$. It follows that $\Delta L_\text{min}^{c}(v^*,\teta^*) \leq \Delta L^*$. On the other hand, should $\Delta L_\text{min}^{c}(v^*,\teta^*) > \Delta L^*$, there should not exist any halo functions contained within the confidence band passing through $(v^*,\teta^*)$. Thus, a two-sided pointwise confidence band can be constructed by finding at each value of $\vmin$, the values of $\teta^*$ around $\teta_{BF}$ which satisfy $\Delta L_\text{min}^{c}(\vmin,\teta^*) \leq \Delta L^*$. For the results presented in \Sec{sec:results}, we plot the contours of $\Delta L^* = 1.0$ and 2.7, which for a chi-squared distribution\footnote{In the limit that $N_{\rm obs}$ is large, Wilk's theorem states that the log-likelihood ratio follows a chi-squared distribution which may not exactly apply with only 3 events.} with one degree of freedom correspond to $68\%$ and $90\%$ CL confidence bands, respectively~\cite{Gelmini:2015voa}. Compatibility of these confidence bands with upper limits can then be assessed at a given CL by determining whether there exists a non-increasing halo function $\teta(\vmin)$ which is entirely contained within a particular band and does not exceed any of the upper limits. A confidence band is said to be excluded if no such halo function can be constructed.


\section{Data Analysis \label{sec:data}}

Here, we present current halo-dependent and halo-independent constraints on the CDMS-II-Si $68\%$ and $90\%$ regions for a variety of elastic and exothermic spin-independent interaction models. We focus explicitly on isospin conserving ($f_n/f_p = 1$),  `Ge-phobic' (defined by the choice of neutron and proton couplings which minimizes scattering in germanium, \ie $f_n/f_p = -0.8$), and `Xe-phobic' models (defined by the choice of neutron and proton couplings which minimizes scattering in xenon, \ie $f_n/f_p = -0.7$). Halo-independent constraints are presented for three representative choices of $m$ and $\delta$, which had been selected in~\cite{Gelmini:2015voa} as parameters in the halo-dependent analyses which appeared to provide good compatibility of the CDMS-II-Si signal and the upper bounds from null searches.

\begin{figure*}
\center
\includegraphics[width=.49\textwidth]{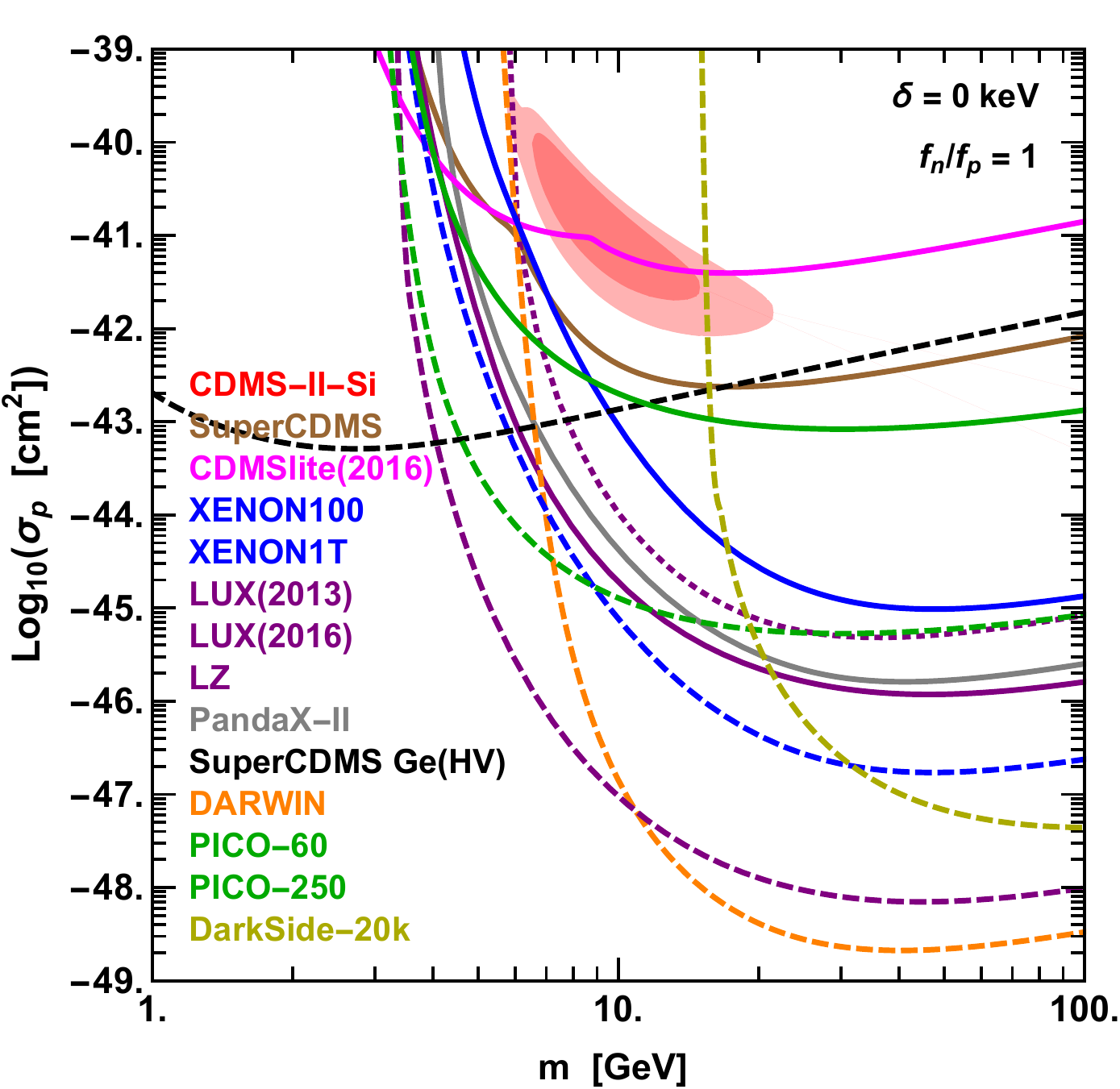}
\caption{\label{fig:delta0_fnfp1} Halo-dependent comparison of CDMS-II-Si $68\%$ (dark red) and $90\%$ (light red) regions with current $90\%$ CL upper limits from SuperCDMS (brown), CDMSlite2016 (magenta), XENON100 (blue, solid), LUX2013 (purple, dotted), LUX2016 (purple, solid), PandaX-II (grey), and PICO-60 (green, solid), for an elastic isospin conserving spin-independent interaction. Also shown are projected discovery limits (dashed) for XENON1T (blue), SuperCDMS SNOLAB Ge HV (black), LZ (purple), DARWIN (orange), DarkSide-20k (yellow), and PICO-250 (green). }
\end{figure*}

Upper limits in this section are presented for the following experiments: SuperCDMS~\cite{Agnese:2014aze}, CDMSlite (2016 result)~\cite{Agnese:2015nto}, XENON100~\cite{Aprile:2016swn}, LUX (2013 result)\footnote{The LUX2013 bound is presented assuming zero observed events. This bound has been shown to be well representative of the true bound~\cite{DelNobile:2013gba}.}~\cite{Akerib:2013tjd}, LUX (2016 result)~\cite{Akerib:2016vxi}, PandaX-II~\cite{Tan:2016zwf}, and PICO-60~\cite{Amole:2017dex}. Also shown are projected bounds for XENON1T~\cite{Diglio:2016stt}, SuperCDMS SNOLAB Ge High-Voltage (which we call SuperCDMS Ge(HV))~\cite{Agnese:2016cpb}, LZ~\cite{Akerib:2015cja,Szydagis:2016few}, DARWIN~\cite{Schumann:2015cpa}, DarkSide-20k~\cite{Agnes:2015ftt,Davini:2016vpd}, and PICO-250~\cite{Pullia:2014vra}. The procedure for constructing the LUX2013 bound was previously outlined in~\cite{DelNobile:2013cta,DelNobile:2013gba,DelNobile:2014eta}. We describe here the process used below to produce the remaining experimental bounds.

\subsection{CDMS-II-Si}
The procedure for analyzing the CDMS-II-Si data follows the procedure outlined in~\cite{DelNobile:2013cta,DelNobile:2013gba,DelNobile:2014eta}. Specifically, we consider the three event signal with energies 8.2, 9.5, and 12.3 keV. CDMS-II-Si had an exposure of 140.2 kg-days and an energy window of $7$ keV to $100$ keV. Using a profile likelihood ratio test, a preference was found for the WIMP+background hypothesis over the background-only hypothesis with a $p-$value of 0.19\%~\cite{Agnese:2013rvf}. We use an $\ER$-dependent efficiency identical to that shown in Fig.~1 of~\cite{Agnese:2013rvf} (solid blue line). Since the energy resolution for silicon in CDMS-II has not been measured, we use a Gaussian resolution function with the energy resolution used for CDMS-II's germanium detectors, taken from in Eq.~1 of~\cite{Ahmed:2009rh}, $\sigma(\Ed) = \sqrt{0.293^2 + 0.056^2 \times \Ed/{\rm keV}}$ keV. To estimate the differential background rate for each observed event, we take the differential background rates from \cite{McCarthy} and normalize each component such that 0.41, 0.13, and 0.08 events are expected from surface events, neutrons, and $^{208}$Pb respectively~\cite{Agnese:2013rvf}. This procedure reproduces the preferred regions shown in Fig.~4 of ~\cite{Agnese:2013rvf}.

\subsection{XENON100}
The XENON100 bound is produced in the manner outlined in~\cite{DelNobile:2013cta}, but using the updated 477 day exposure~\cite{Aprile:2016swn}. This procedure accurately reproduces the bound shown in Fig.~11 of \cite{Aprile:2016swn}.

\subsection{CDMSlite 2016}

The CDMSlite bound (hereby CDMSlite2016) is constructed using results from the recently reported 70.1 kg-day exposure. The detector efficiency and quenching factor are taken from Fig.~1 and Eq.~3 of \cite{Agnese:2015nto}, respectively. The energy of detected events is read off the inset in Fig.~3 in~\cite{Agnese:2015nto}, but only between detected energies of 0.36 and 1.04 keVee, and the maximum gap method is then applied. This procedure reproduces the published bound.

\begin{figure*}
\center
\includegraphics[width=.49\textwidth]{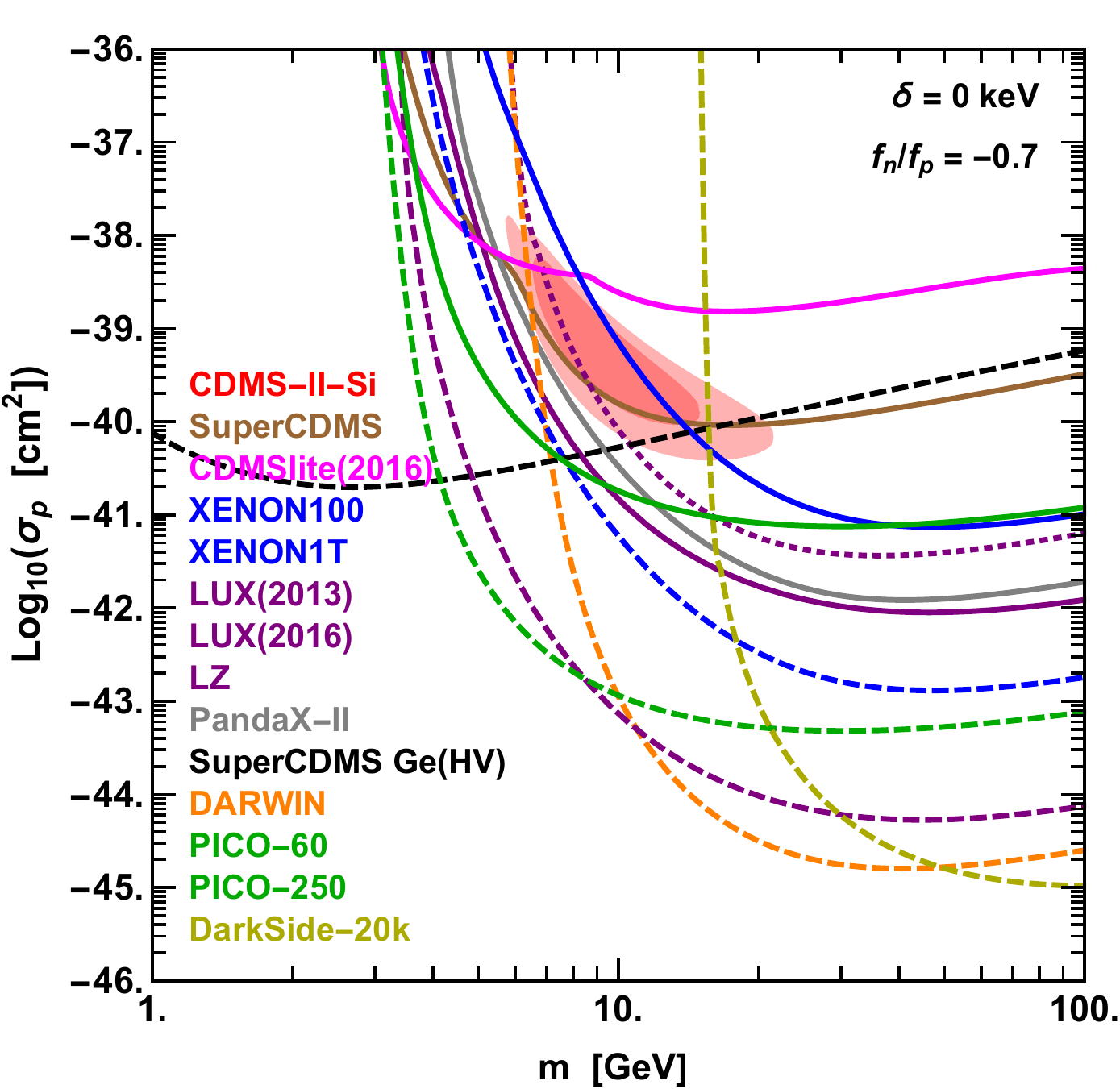}
\includegraphics[width=.49\textwidth]{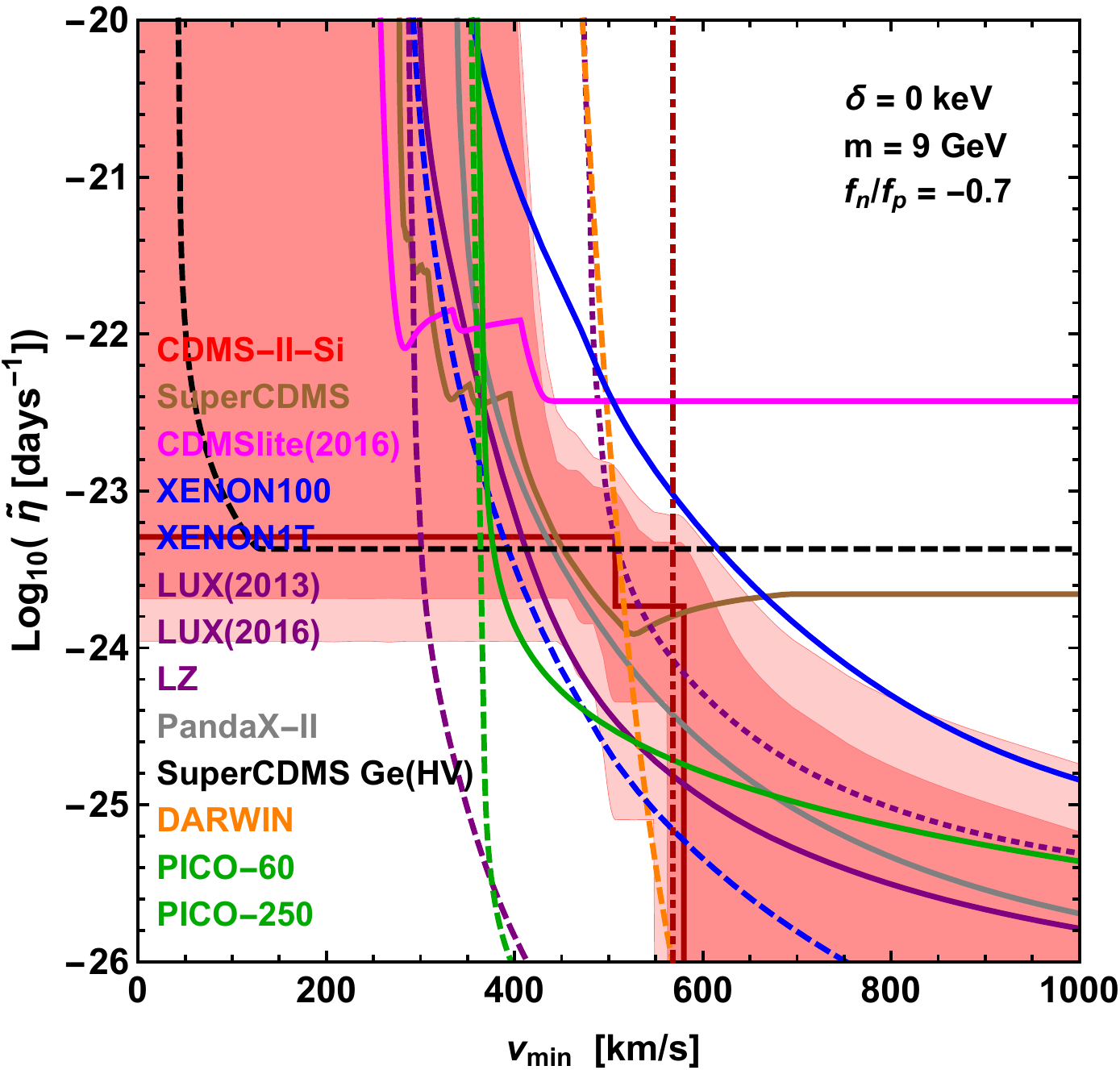}
\caption{\label{fig:delta0_fnfp_neg} (Left) Halo-dependent analysis and (Right) halo-independent analysis for $m = 9$ GeV, assessing the compatibility of the CDMS-II-Si $68\%$ and $90\%$ CL regions (shown in darker and lighter red) with the $90\%$ CL upper limits and projected sensitivities of other experiments, for an elastic spin-independent contact interaction with $f_n/f_p = -0.7$. We include SuperCDMS (brown), CDMSlite (magenta), XENON100 (blue, solid), LUX2013 (purple, dotted), LUX2016 (purple, solid), PandaX-II (grey), and PICO-60 (green, solid) upper limits, and the projected sensitivities (dashed lines) of XENON1T (blue), SuperCDMS Ge(HV) (black), LZ (purple), DARWIN (orange), DarkSide-20k (yellow), and PICO-250 (green). Also shown is the best-fit halo function $\teta_{BF}$ to the CDMS-II-Si data (dark red step function) and the $\vmin$ value corresponding to the event with the largest observed recoil energy, assuming $\ER=\Ed=12.3$ keV (vertical dot dashed dark red line). }
\end{figure*}

\subsection{LUX 2016}

The LUX bound is computed by using the complete LUX exposure (approximately $4.47\times10^{4}$ kg-days). The efficiency and fractional resolution as functions of $\ER$ are extracted from Fig.~2 (black solid line) and Fig.~5 of \cite{Akerib:2016vxi}, respectively. The bound is obtained by determining the cross section required to produce a total of 3.2 events. As mentioned in~\cite{Akerib:2016vxi} this procedure reproduces the $90\%$ CL combined LUX exclusion limit. 

\subsection{PandaX-II}

The constraint for PandaX-II is based on the $3.3\times10^4$ kg-day run data published in 2016. To reproduce the published bound, the nuclear recoil efficiency function is taken from Fig.~2 of~\cite{Tan:2016zwf} (black line), and the recoil energies of the three observed events are read off Figs.~4 and 14 of \cite{Tan:2016zwf}). Applying the maximum gap method~\cite{Yellin:2002xd} yields a bound that reproduces well the published bound for $m \lesssim 30$ GeV, and is slightly stronger at larger masses by a factor of $\lesssim 1.5$.

\begin{figure*}
\center
\includegraphics[width=.4\textwidth]{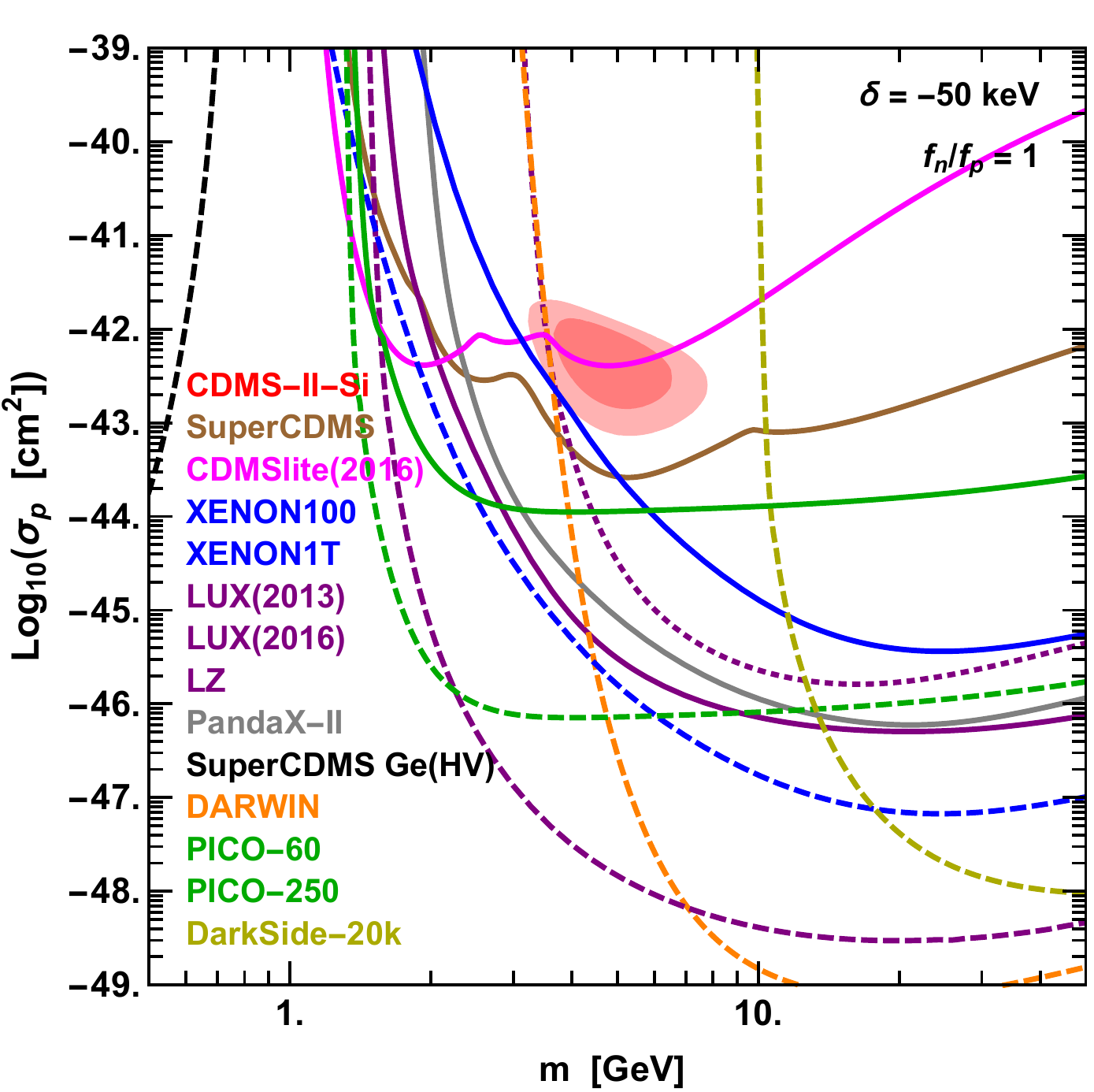}
\includegraphics[width=.4\textwidth]{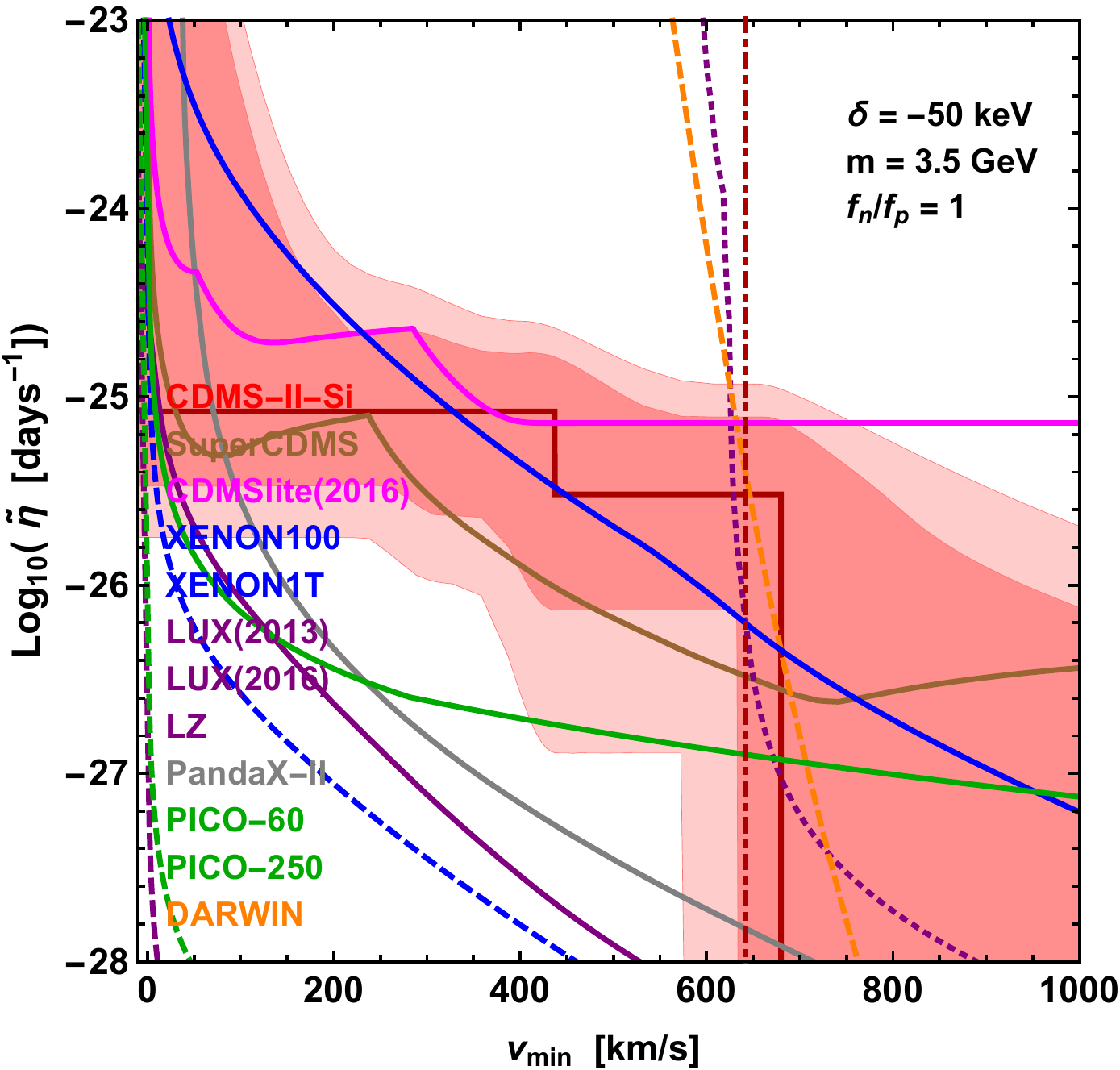}
\includegraphics[width=.4\textwidth]{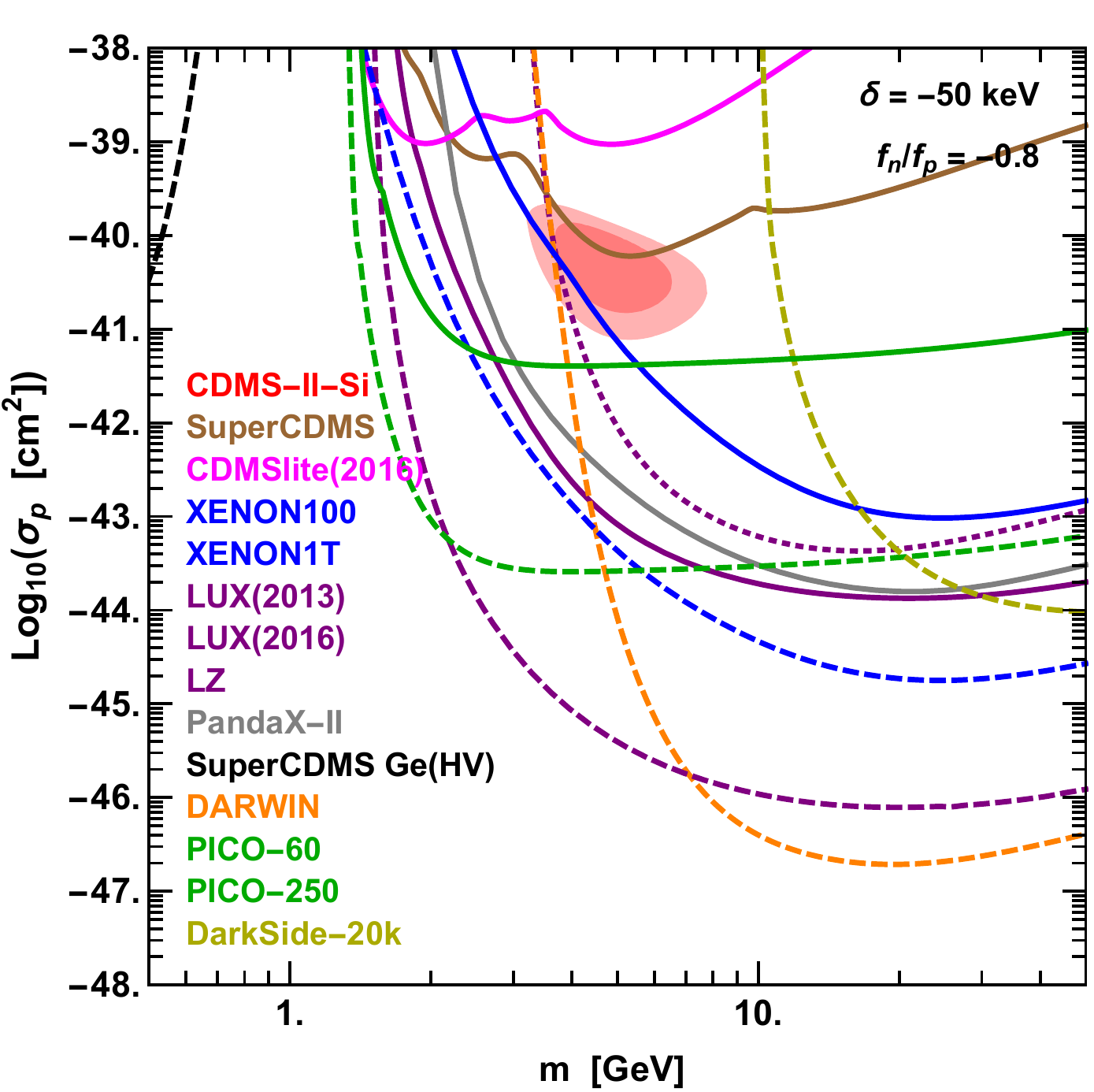}
\includegraphics[width=.4\textwidth]{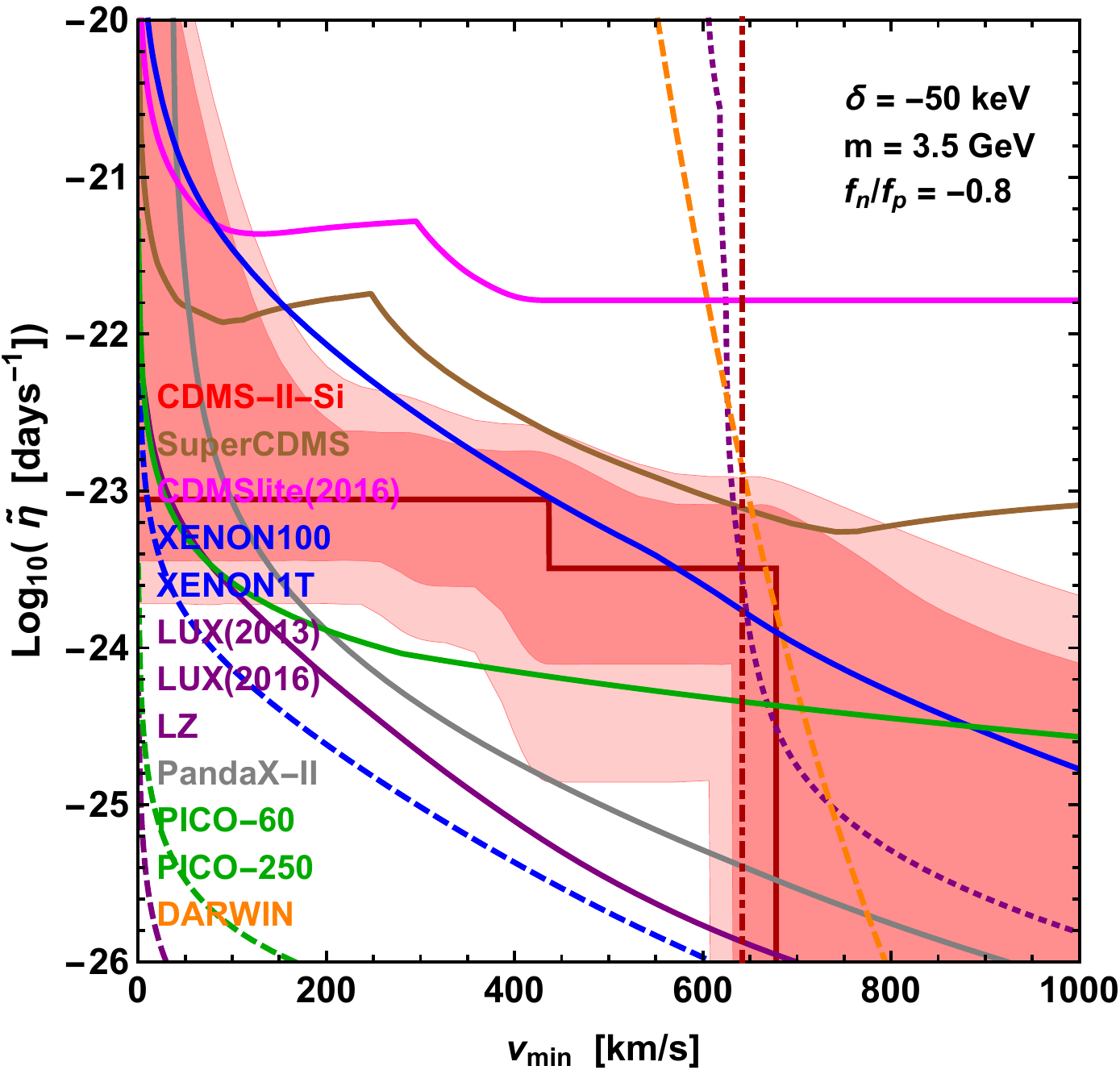}
\includegraphics[width=.4\textwidth]{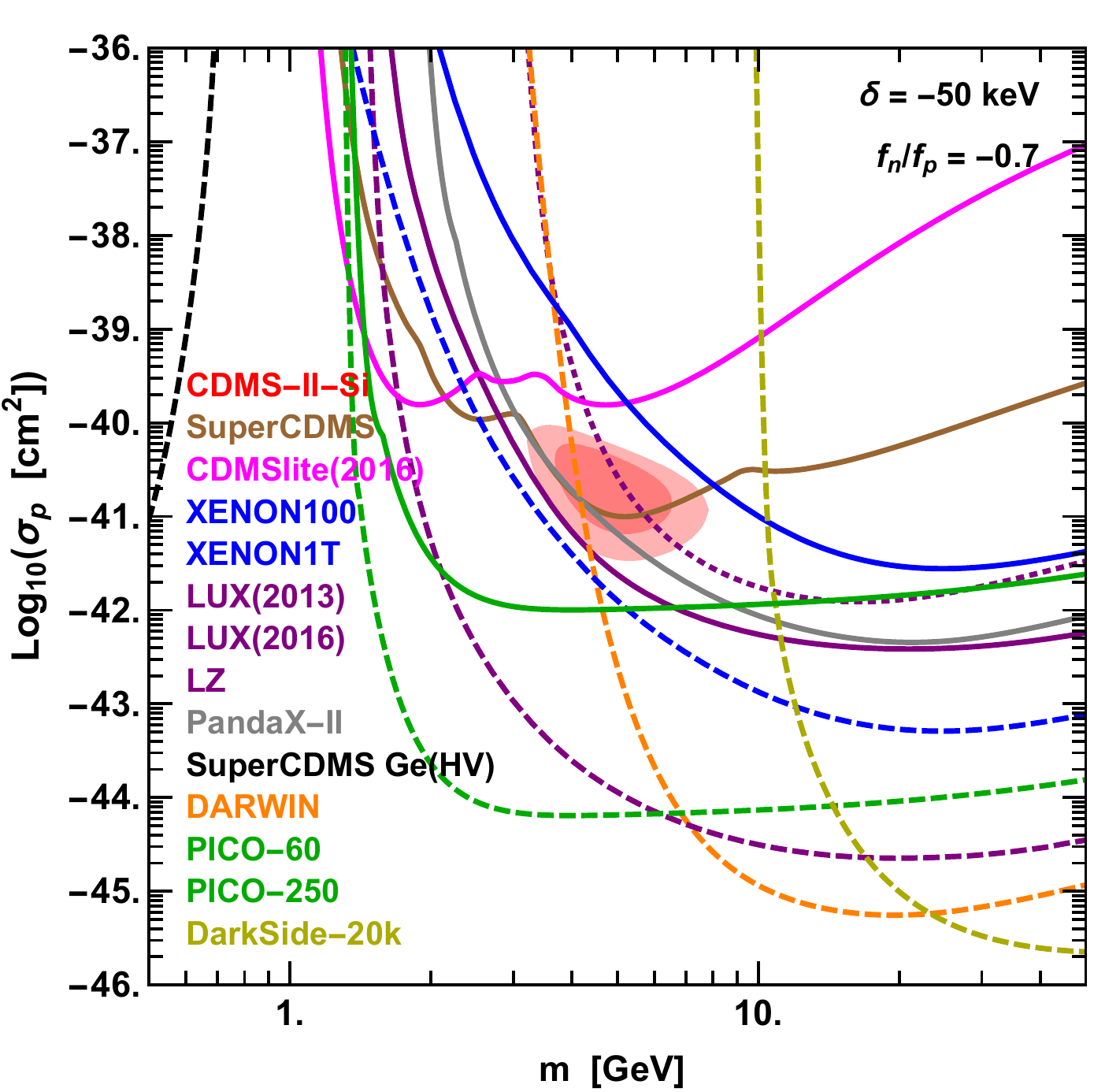}
\includegraphics[width=.4\textwidth]{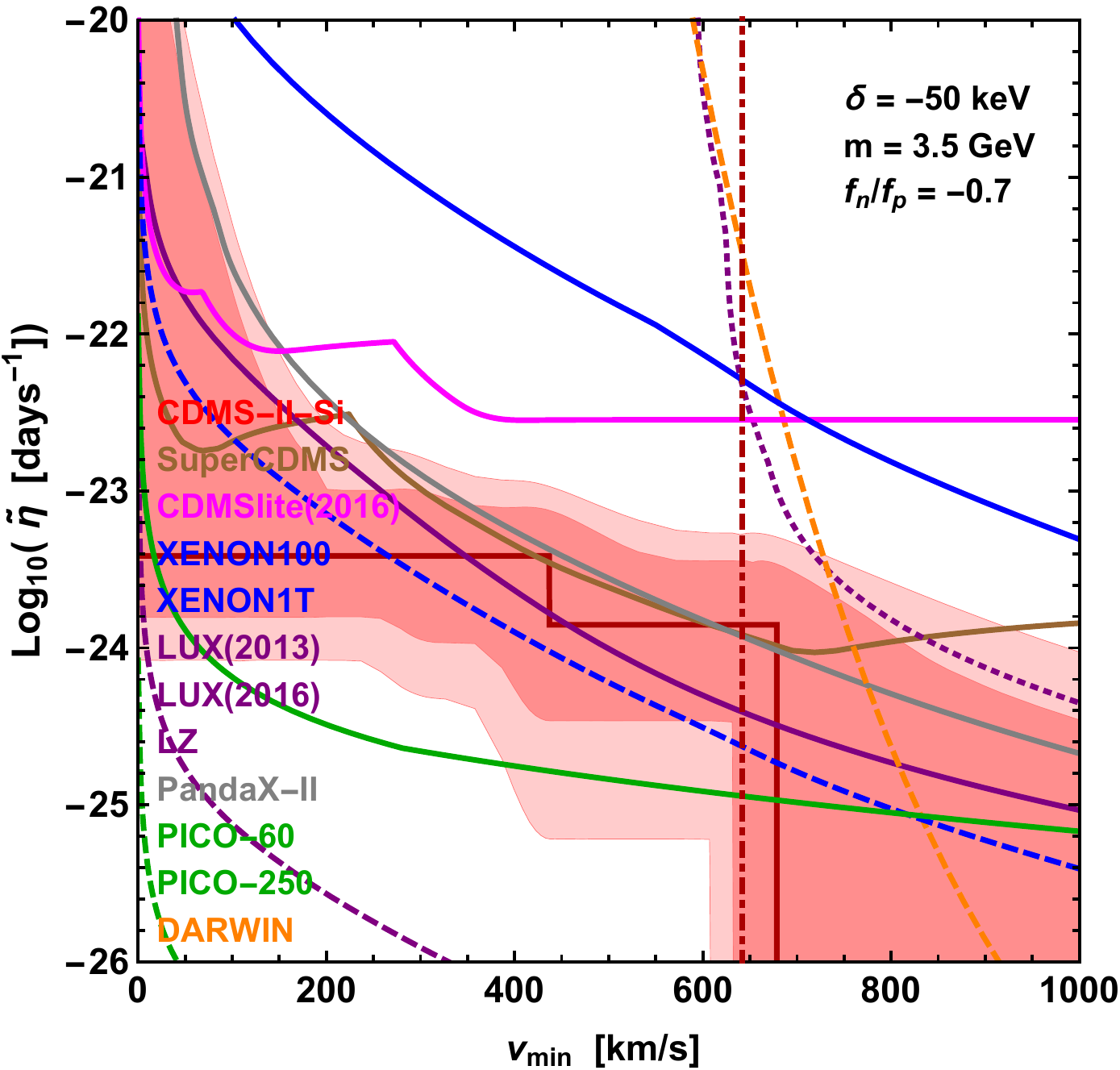}
\caption{\label{fig:delta_neg50_fnfp1} 
(Left) Halo-dependent analysis and (Right) halo-independent analysis for $m = 3.5$ GeV, assessing the compatibility of the CDMS-II-Si $68\%$ and $90\%$ CL regions (shown in darker and lighter red) with the $90\%$ CL upper limits and projected sensitivities of other experiments, for an exothermic spin-independent contact interaction with $\delta = -50$ keV. Results are shown for isospin conserving couplings (top), `Ge-phobic' couplings (middle), and `Xe-phobic' couplings (bottom). Experiments included are identical to those shown in \Fig{fig:delta0_fnfp_neg}. }
\end{figure*}

\subsection{PICO-60}\label{subsec:pico60}

The constraint for PICO-60 is based on the recent $1167$ kg-day run of $C_3F_8$~\cite{Amole:2017dex}. Here, we restrict our attention to scattering off fluorine, as this element accounts for $\simeq 80\%$ of the target mass and has a lower threshold than carbon (after considering the bubble nucleation efficiency in Fig.~4 of~\cite{Amole:2015lsj}). PICO-60 is run at a thermodynamic threshold of $3.3$ keV, however this threshold does not correspond to the threshold recoil energy in fluorine required to nucleate a bubble. We take this threshold to be $6$ keV using the efficiency function shown in Fig.~4 of \cite{Amole:2015lsj} for a 3.2 keV thermodynamic threshold (although this is only determined for a $5$ GeV WIMP with a spin-independent interaction). Using Poisson statistics with zero observed events and zero expected background, we find this threshold perfectly reproduces the published bound~\cite{Amole:2017dex}.

\subsection{XENON1T}

The projected bound for XENON1T~\cite{Diglio:2016stt} is computed assuming a 2 ton-year exposure, a flat efficiency of $0.4$, and an effective light yield, a low-energy threshold, and an energy resolution equal to those used in the XENON100 analysis of~\cite{DelNobile:2013cta}. This procedure produces a sensitivity limit consistent with the $\pm 1\sigma$ confidence intervals of the 2 ton-year sensitivity limit shown in Fig.~8 of \cite{Diglio:2016stt}.

\subsection{SuperCDMS SNOLAB Ge(HV)}

SuperCDMS plans to operate the next generation of their experiment at SNOLAB beginning in 2020; the discovery limits produced here are based on the recent projected sensitivity for their high-voltage germanium, Ge(HV), detectors. Specifically, we assume 8 Ge(HV) detectors, each with an exposure of 44 kg-days. We also assume perfect efficiency in the energy range $0.04$ keV (taken from Table VIII of \cite{Agnese:2016cpb}) to $2$ keV (taken to be consistent with the energy range suggested in the caption of Table V of~\cite{Agnese:2016cpb}), perfect energy resolution, an ionization yield given by Lindhard theory (with parameters taken from \cite{Scholz:2016qos}), and zero observed events. Using the maximum gap method (which coincides with using a Poisson likelihood in this case) we obtained a $90\%$ CL limit very similar to the Ge(HV) limit shown in Fig. 8 of \cite{Agnese:2016cpb}. SuperCDMS also plans to run a high-voltage silicon detector which is not included here because its projected sensitivity is inferior across most of the parameter space. Also note that if the energy ranges of the HV detectors could be extended to energies beyond $2$ keV, these experiments could gain sensitivity to the exothermic models considered here.

\subsection{LZ}
 
The projected sensitivity for LZ is produced using the same energy resolution and efficiency function used in the LUX2016 analysis, and assuming a total exposure of $15.33$ ton-years (\ie a $5.6$ ton fiducial volume with 1000 live-days)~\cite{Akerib:2015cja,Szydagis:2016few}. We then apply the maximum gap method, under the assumption of zero observed events, with which we reproduce a sensitivity limit comparable to that shown in Fig.4 of \cite{Szydagis:2016few}.

\subsection{DARWIN}

The projected sensitivity limit for DARWIN is based on the design presented in~\cite{Schumann:2015cpa}, for a liquid xenon experiment with a 200 ton-year exposure. Following~\cite{Schumann:2015cpa}, we consider an energy range of $5$ keV to $20.5$ keV and a constant detection efficiency of $30\%$. We approximate the energy resolution as a Gaussian with $\sigma = \ER \times 0.15$, which is roughly consistent with Fig.~1 of \cite{Schumann:2015cpa}. Assuming zero observed events, the bound is obtained using the maximum gap method. This procedure is found to produce a sensitivity limit in strong agreement with that shown in Fig.~7 of \cite{Schumann:2015cpa}.

\subsection{DarkSide-20k}
The projected sensitivity for DarkSide-20k is produced assuming a flat nuclear recoil efficiency of $0.7$ between energies $40$ keV and $240$ keV (and zero elsewhere), a 60 ton-year exposure (\ie a 20 ton fiducial volume run for 3 years), and by applying the maximum gap method with zero observed events~\cite{Agnes:2015ftt,Davini:2016vpd}. DarkSide-20k is not sensitive for the nuclear recoils imparted to argon nuclei by the particular candidates in our halo-independent analyses (we show $\vmin$ $\leq 1000$ km/s), thus no DarkSide-20k bounds appear in the halo-independent plots.

\subsection{PICO-250}
The projected sensitivity for PICO-250 is produced assuming perfect detection efficiency for energies above $6$ keV (see \Sec{subsec:pico60}), a $250$ kg fiducial volume, a $2$ year runtime (or alternatively, a $500$ kg fiducial volume run for one year), and by using Poisson statistics with zero observed events and zero expected background~\cite{Pullia:2014vra,Amole:2017dex}. As in \Sec{subsec:pico60}, we only consider scattering off fluorine.

\section{Results \label{sec:results}}

\begin{figure*}
\center
\includegraphics[width=.4\textwidth]{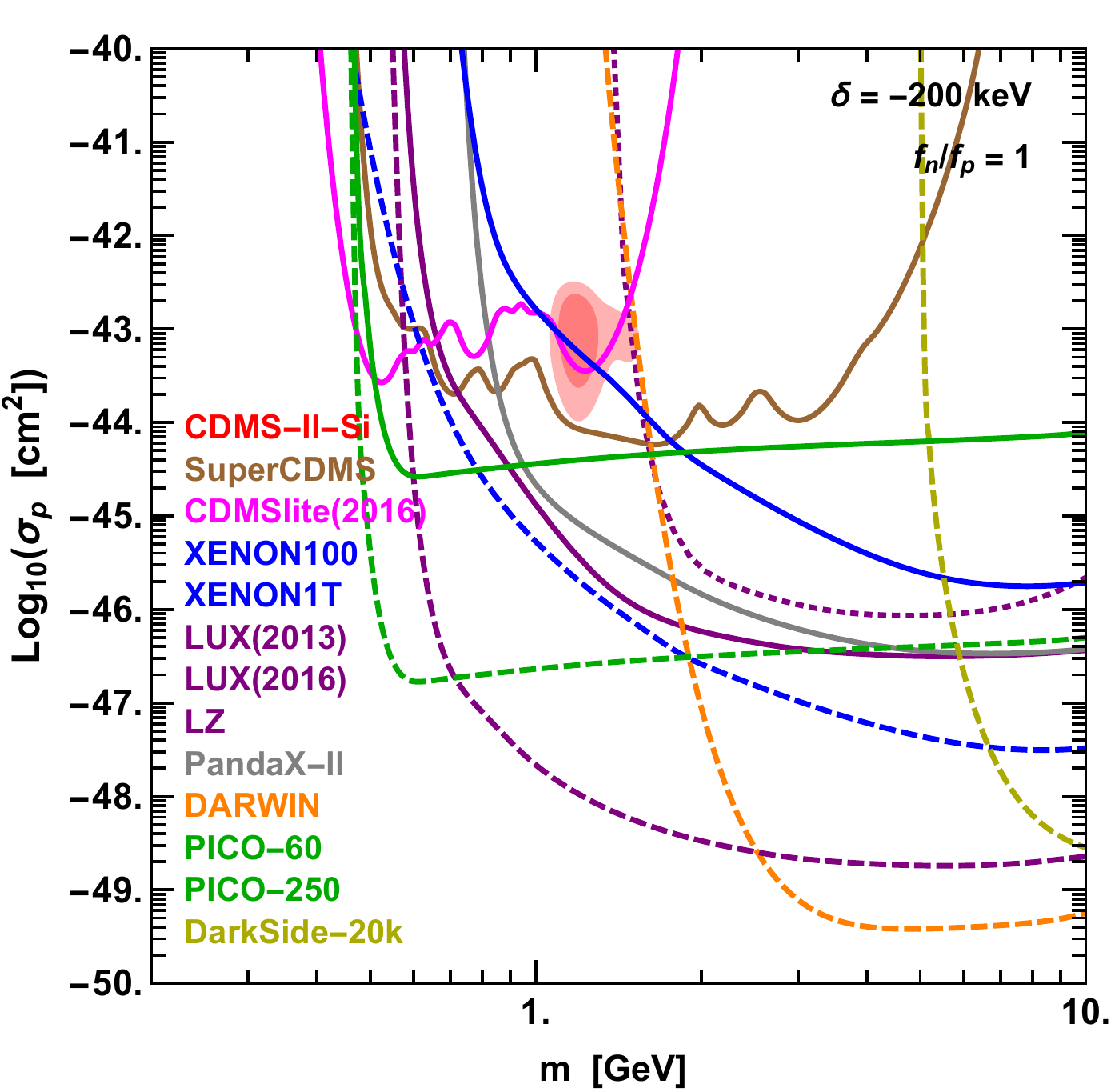}
\includegraphics[width=.4\textwidth]{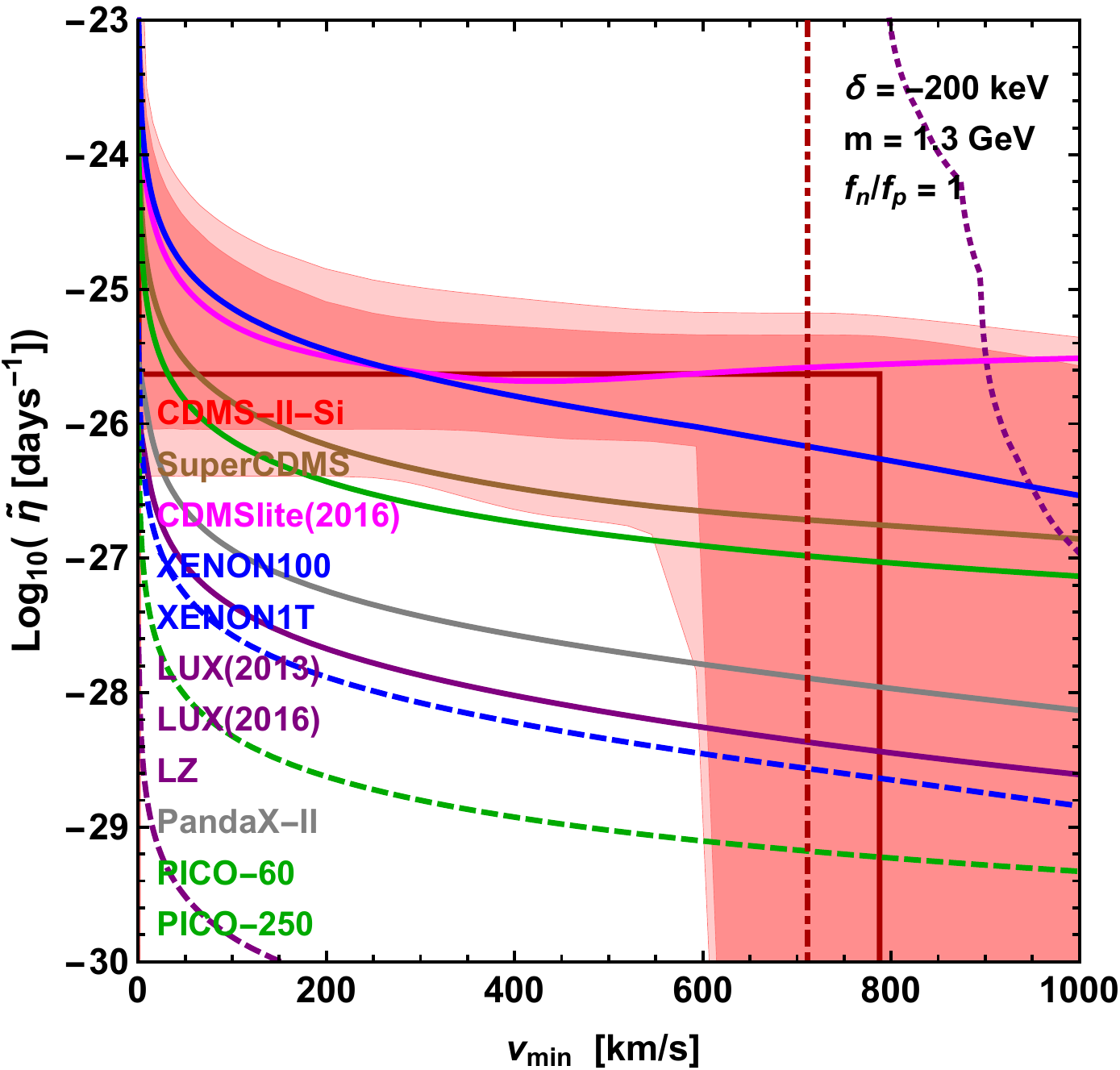}
\includegraphics[width=.4\textwidth]{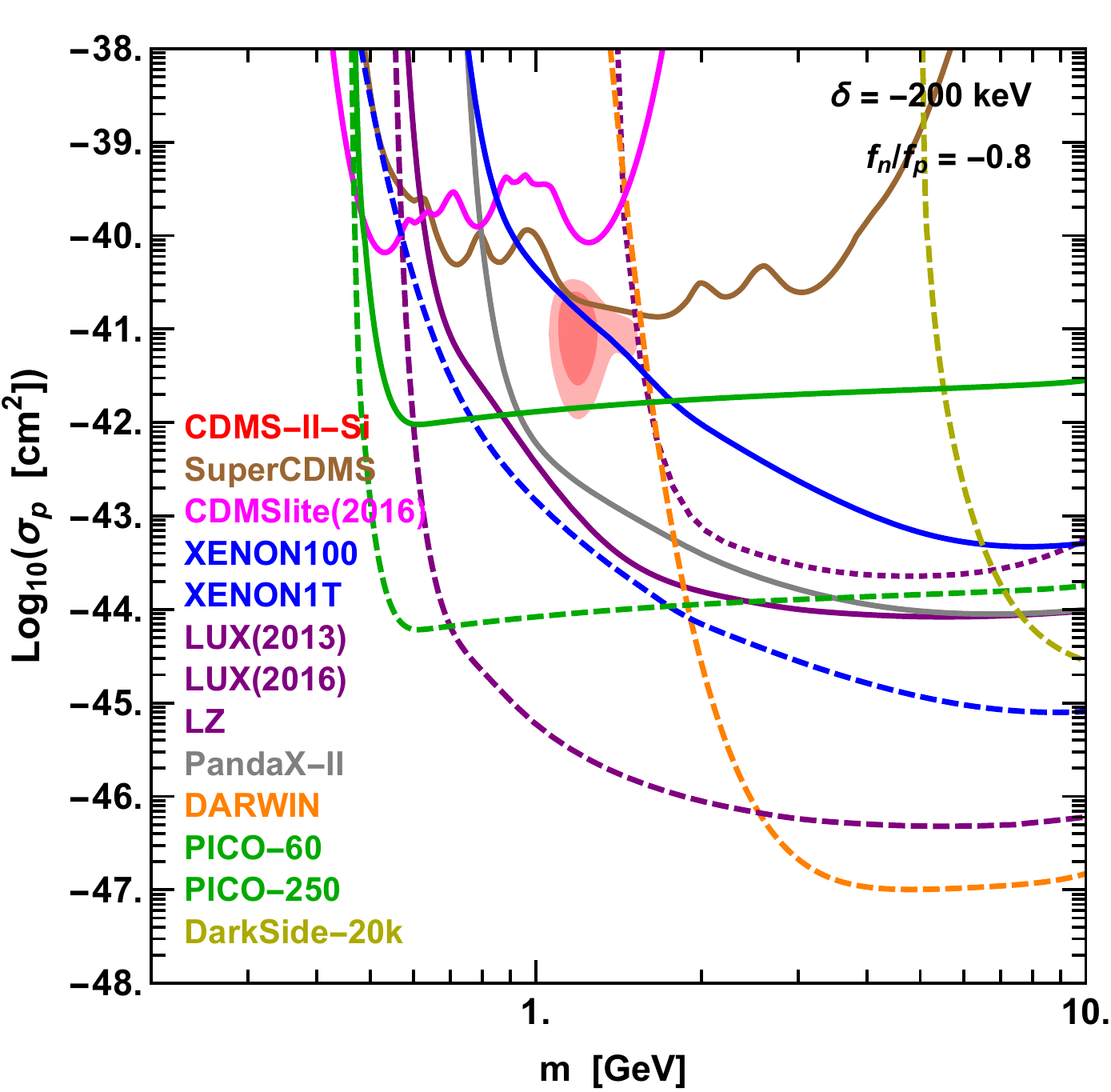}
\includegraphics[width=.4\textwidth]{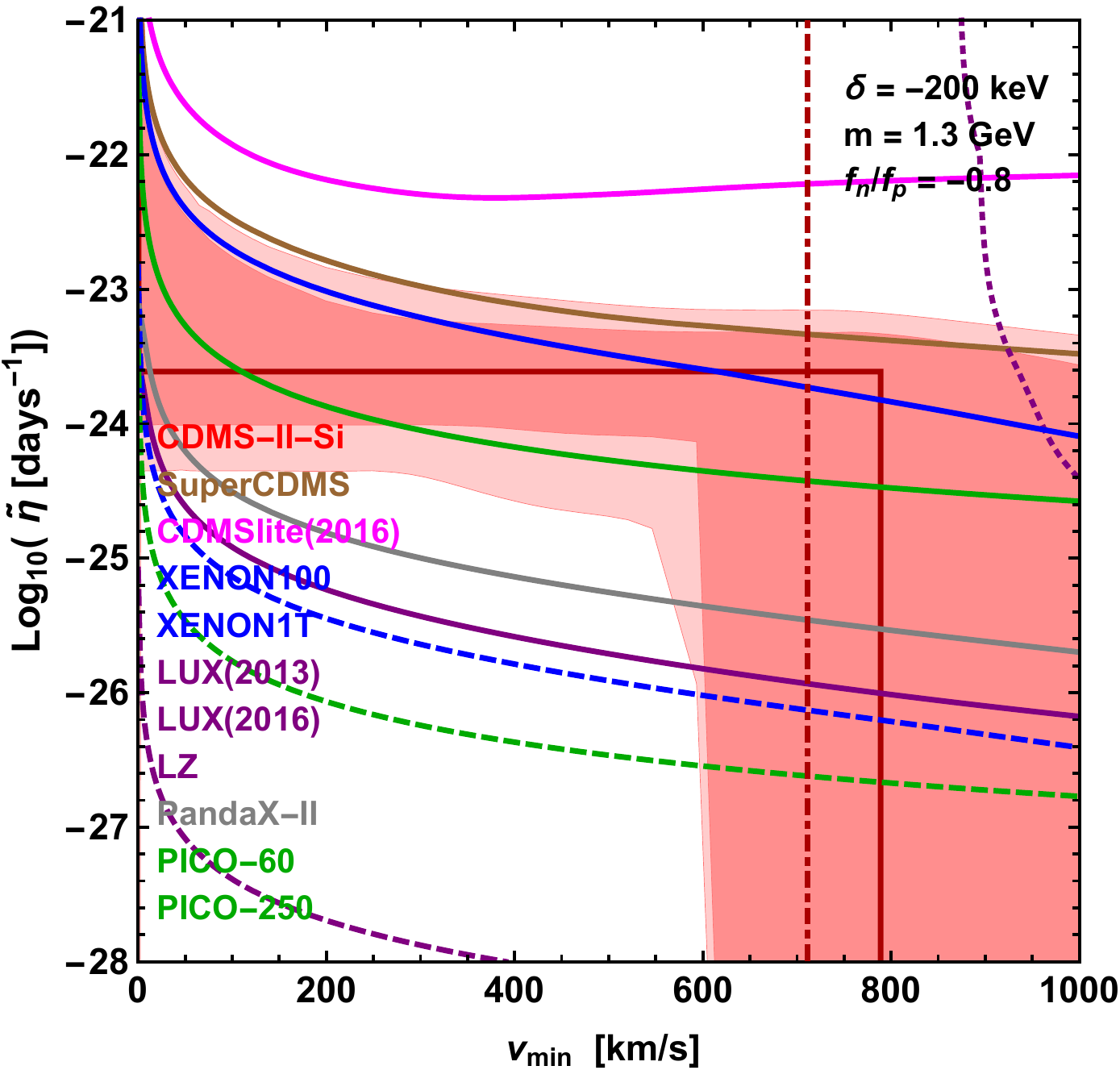}
\includegraphics[width=.4\textwidth]{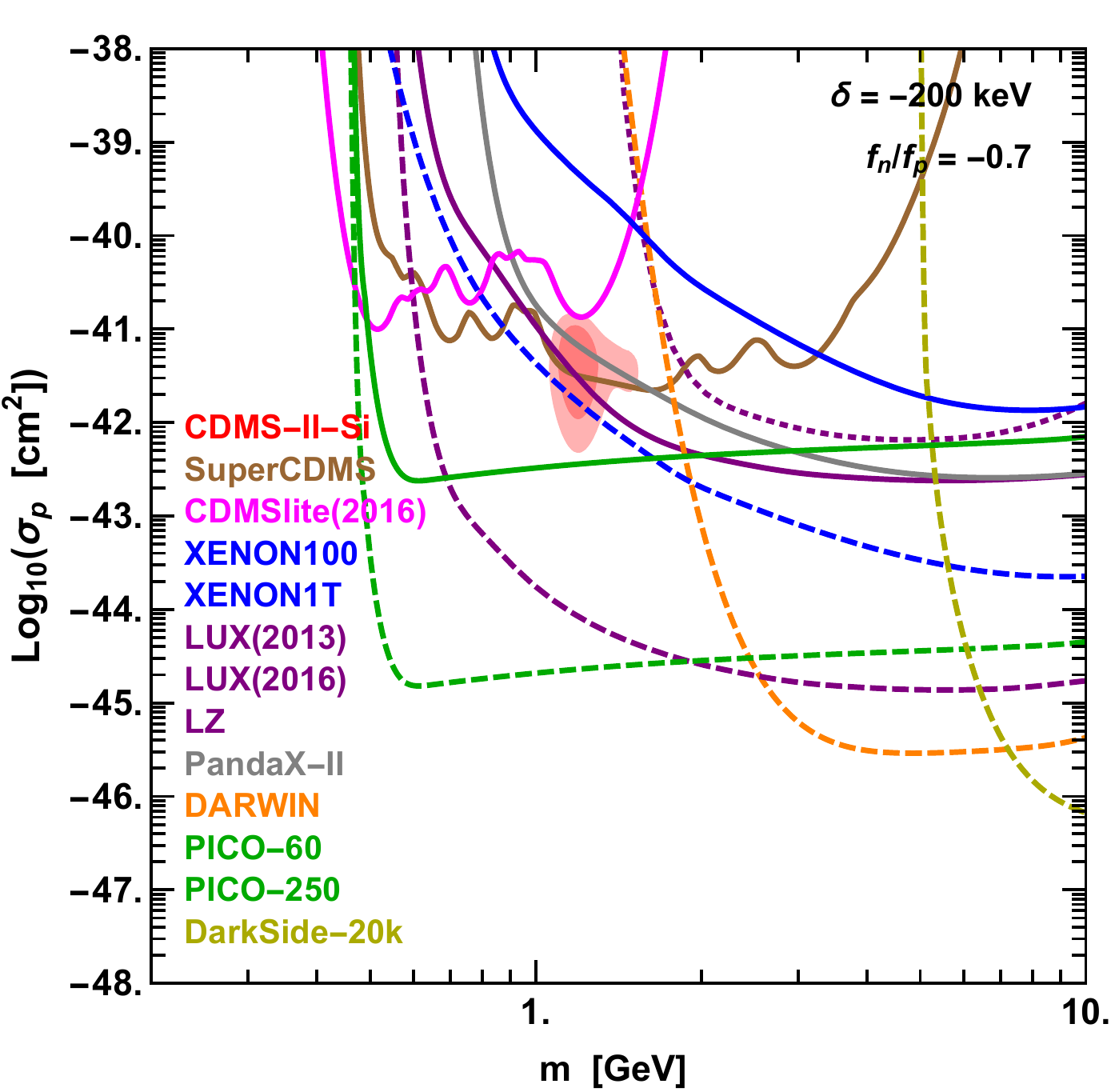}
\includegraphics[width=.4\textwidth]{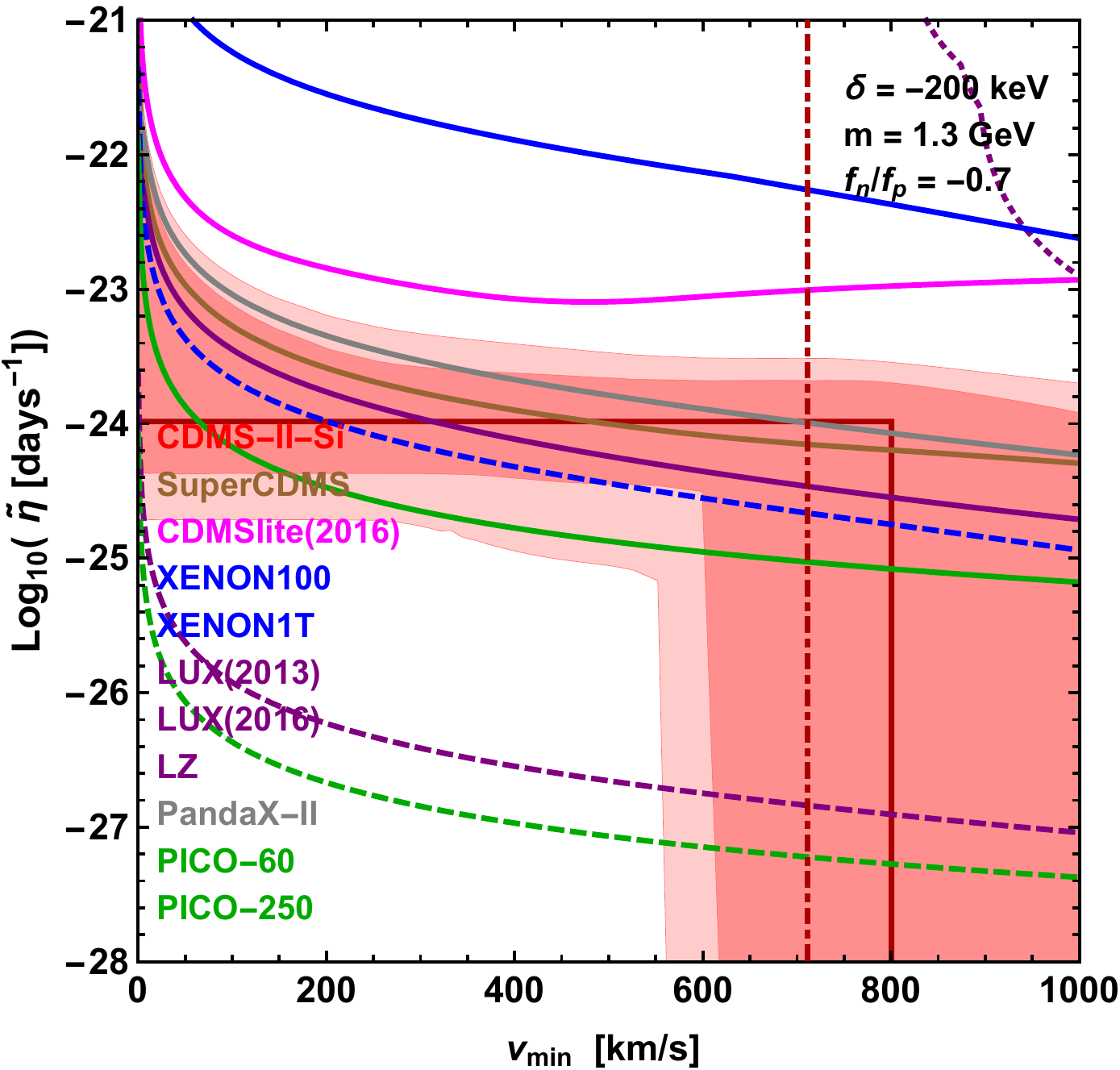}
\caption{\label{fig:delta_neg200_fnfp1} Same as \Fig{fig:delta_neg50_fnfp1}, but for $\delta = -200$ keV and $m = 1.3$ GeV (halo-independent analyses only). The SuperCDMS Ge(HV) discovery limit is not shown as it cannot probe the WIMP candidates shown. }
\end{figure*}

For the purpose of providing context, we begin by plotting in \Fig{fig:delta0_fnfp1} a comparison of the $68\%$ and $90\%$ CDMS-II-Si regions with the current and projected $90\%$ CL limits of other experiments, assuming the conventional elastic spin-independent contact interaction with isospin conserving couplings. Null results from LUX2013 and SuperCDMS have excluded this model at the $90\%$ CL in both halo-dependent and halo-independent analyses (there exist small discrepancies in the preferred CDMS-II-Si regions of~\cite{Gelmini:2015voa} and those presented below, a mistake that arose because the factor of 2 in the definition of $L[\teta]$ was missing in~\cite{Gelmini:2015voa})~\cite{DelNobile:2013gba,Gelmini:2015voa}. 

We present in \Fig{fig:delta0_fnfp_neg} a halo-dependent (left) and halo-independent (right) analysis of an elastic spin-independent contact interaction with `Xe-phobic' couplings (\ie $f_n/f_p = -0.7$). In the halo-dependent analysis, the $90\%$ CL CDMS-II-Si region is excluded by the $90\%$ CL upper limits of LUX2016, PandaX-II, and PICO-60. This is consistent with the results of~\cite{Geng:2016uqt}. In the halo-independent analysis, the upper limit of PandaX-II does not entirely exclude the $68\%$ CL CDMS-II-Si region, the LUX2016 limit only marginally excludes the  $90\%$ CL CDMS-II-Si region, and only the very recent PICO-60 $90\%$ CL bound definitively excludes $90\%$ CL CDMS-II-Si region. This is shown for $m = 9$ GeV, but other choices of masses lead to similar results. In the halo-independent analysis, we also show the $\vmin$ value corresponding to the energy of the event with the largest observed energy, assuming $\ER = \Ed = 12.3$ keV (shown with vertical dot dashed dark red line). Had our analysis of the CDMS-II-Si data assumed a perfect energy resolution, the location of the highest step of the best-fit $\teta$ would identically correspond to this value of $\vmin$; with finite energy resolution, the locations of the steps of the best-fit $\teta$ function occur at slightly larger values of $\vmin$. For highly exothermic models, it becomes important to verify that the dark matter speeds capable of producing such recoils are physical, \ie they do not exceed the galactic escape velocity, which for the Standard Halo Model is $v_{\rm esc} \simeq 765$ km/s (in the lab frame).

\begin{figure*}
\center
\includegraphics[width=.49\textwidth]{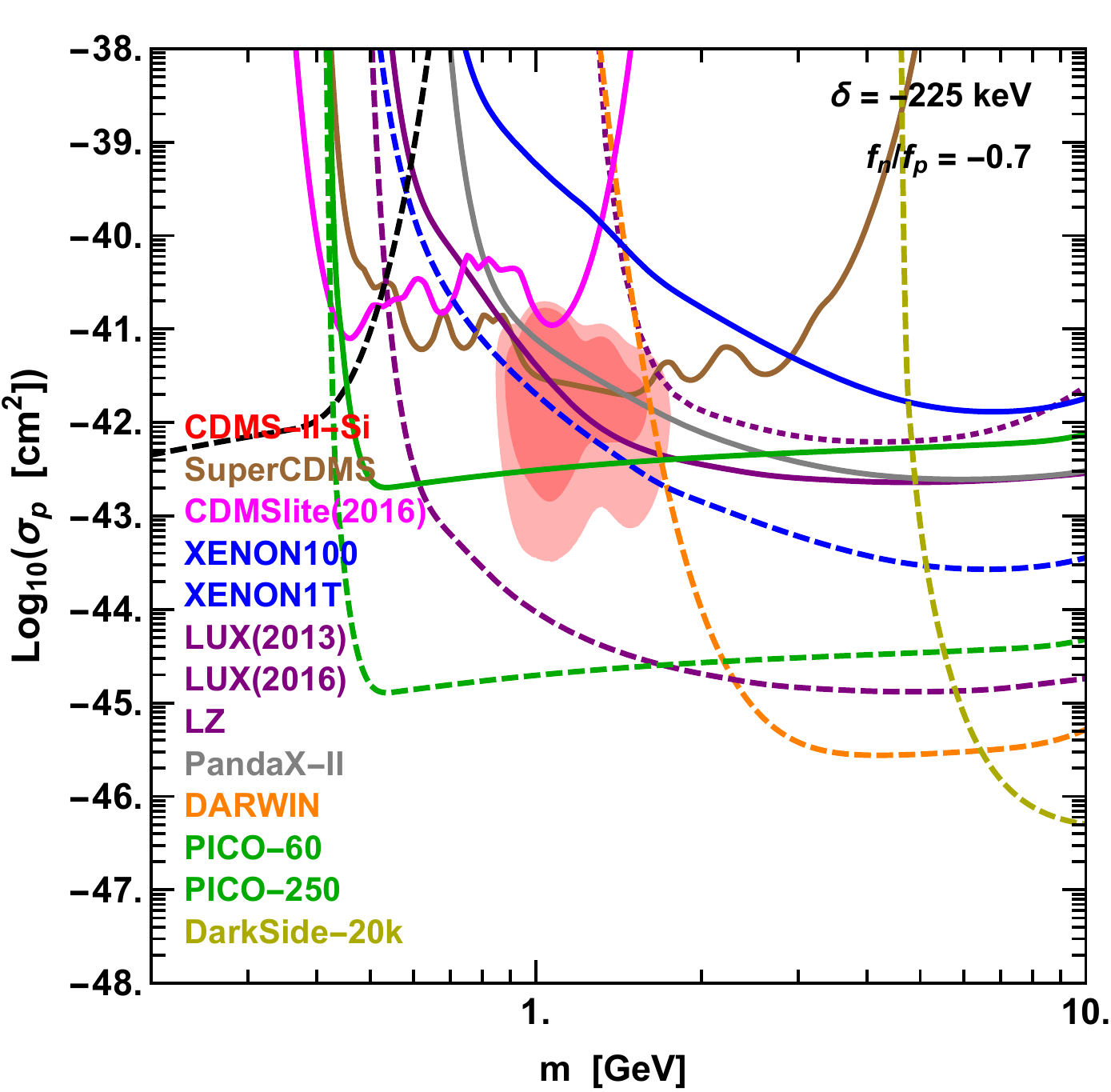}
\includegraphics[width=.47\textwidth]{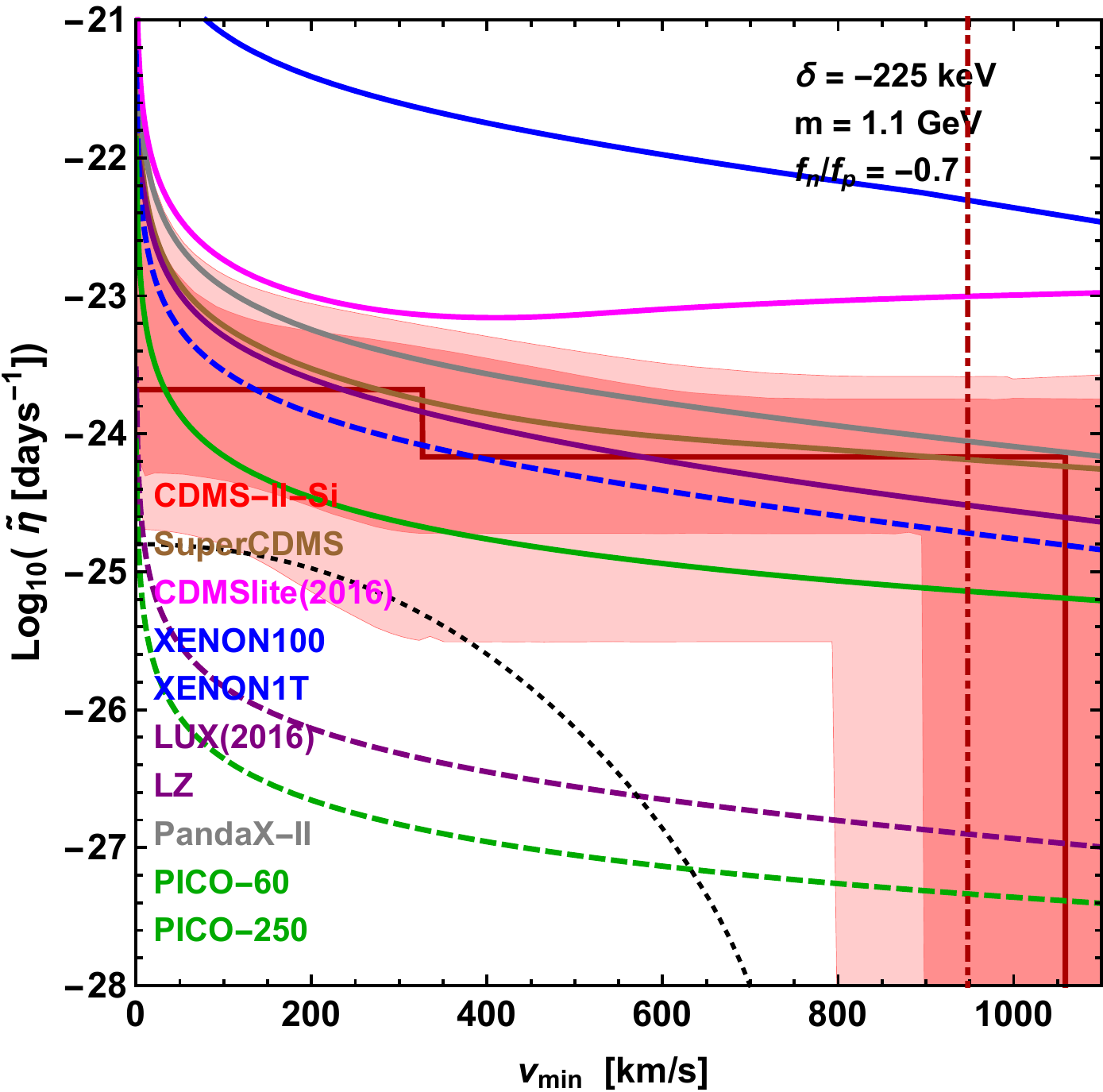}
\caption{\label{fig:delta_neg225_fnfp_neg07} Same as \Fig{fig:delta0_fnfp_neg} but for $\delta = -225$ keV. The halo-independent analysis is shown for $m = 1.1$ GeV. The dotted black line in the halo-independent analysis shows the SHM $\tilde{\eta}(\vmin)$ function for $\sigma_p = 2 \times 10^{-43}\rm{cm}^2$, a value included in the 68\% CL region in the halo-dependent analysis.}
\end{figure*}

In Figs.~\ref{fig:delta_neg50_fnfp1}, \ref{fig:delta_neg200_fnfp1} and~\ref{fig:delta_neg225_fnfp_neg07} we plot halo-dependent (left) and halo-independent (right) analyses of exothermic spin-independent contact interactions with $\delta = -50$ keV, $\delta = -200$ keV, and $\delta = -225$ keV respectively. In Figs.~\ref{fig:delta_neg50_fnfp1} and \ref{fig:delta_neg200_fnfp1} results are shown for isospin conserving (top), `Ge-phobic' (middle), and `Xe-phobic' (bottom) models. The halo-dependent analyses in \Fig{fig:delta_neg50_fnfp1} show that the present $90\%$ CL limits reject the $68\%$ and $90\%$ CL CDMS-II-Si regions. The \Fig{fig:delta_neg50_fnfp1} halo-independent analyses, shown for $m = 3.5$ GeV, illustrate that the CDMS-II-Si $90\%$ CL region for a `Xe-phobic' interaction with $\delta = - 50$ keV is only excluded by the recent PICO-60, and not by the PandaX-II or LUX limits. Note that the $2$ keV upper cutoff imposed on the recoil energy in the SuperCDMS Ge(HV) data analysis implies that this experiment only tests very light exothermic candidates, and does not probe the CDMS-II-Si regions. Similarly, DARWIN's relatively large low energy threshold prevents this experiment from probing the WIMP candidate presented in the halo-independent analysis. This is a consequence of only showing WIMP speeds less than $1000$ km/s.

\begin{figure*}
\center
\includegraphics[width=.49\textwidth]{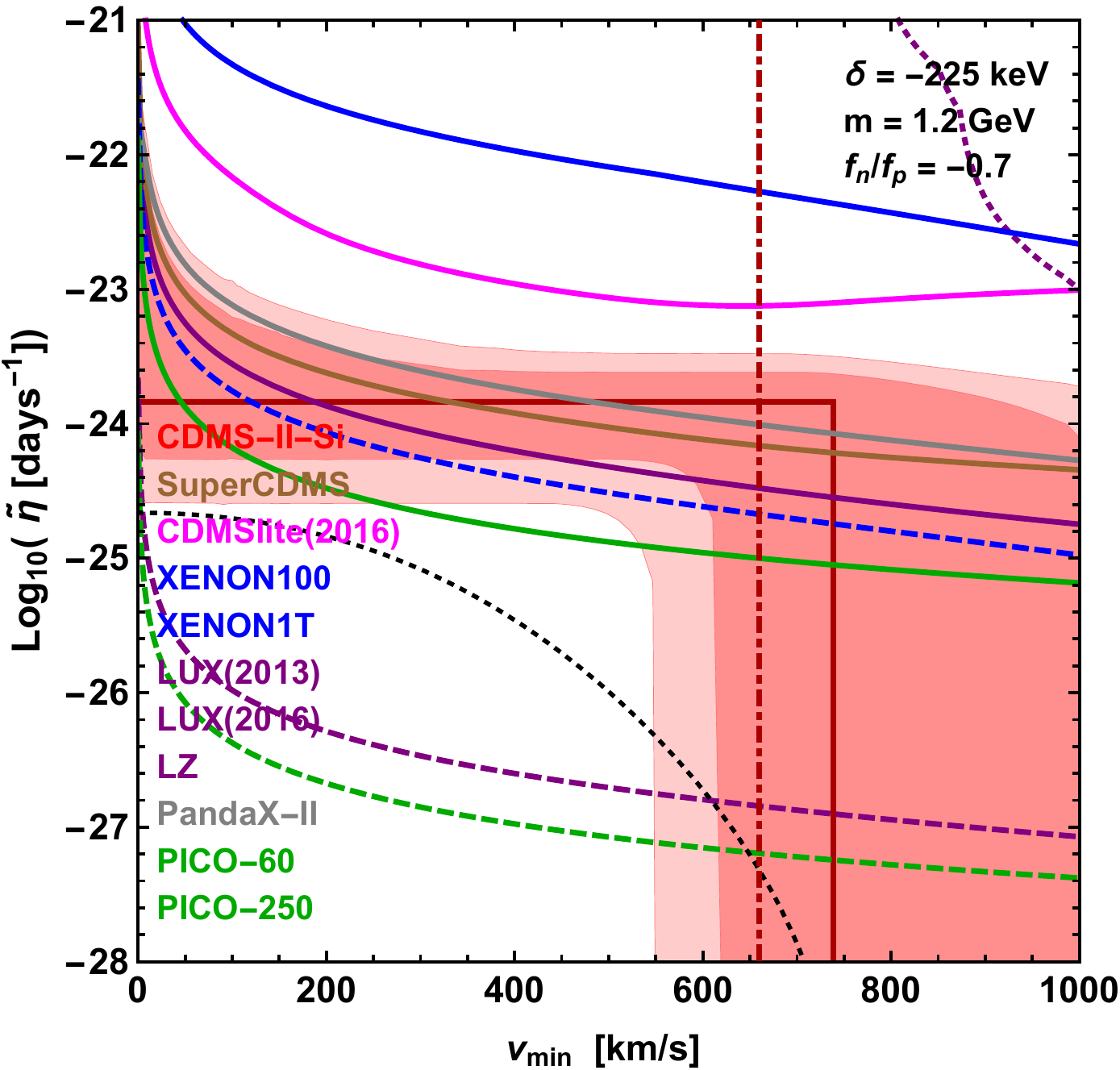}
\caption{\label{fig:delta_neg225_fnfp_neg07_omass}Same as right panel of \Fig{fig:delta_neg225_fnfp_neg07} but for $m = 1.2$ GeV. The dotted black line in the halo-independent analysis shows the SHM $\tilde{\eta}(\vmin)$ function for $\sigma_p = 3 \times 10^{-43}\rm{cm}^2$, a value included in the 90\% CL region in the halo-dependent analysis. }
\end{figure*}

The results shown in \Fig{fig:delta_neg200_fnfp1} are similar to those in \Fig{fig:delta_neg50_fnfp1}, except that in the halo-independent analyses (shown for $m = 1.3$ GeV), the  $90\%$ CL CDMS-II-Si region for the `Xe-phobic' interaction with $\delta = -200$ keV is no longer ruled out for a small set of halo functions which deviate considerably from the SHM. It would seem that increasingly exothermic scattering kinematics (\ie more negative values of $\delta$) may alleviate the tension between the dark matter interpretation of CDMS-II-Si and the null results of other experiments. This is not the case, however, as increasingly negative values of $\delta$ decrease the range of recoil energies that can be imparted by WIMPs (see \Eq{eq:Ebranch}). This implies that highly exothermic candidates traveling at speeds less than the galactic escape velocity may not be able to account for all three events observed by CDMS-II-Si (as illustrated in Fig.~1  of ~\cite{Gelmini:2014psa}). While the largest step in the best-fit $\teta$ function in \Fig{fig:delta_neg200_fnfp1} does lie above what is conventionally taken to be the galactic escape velocity, $\sim765$ km/s in the lab frame, the $\vmin$ value corresponding to the $12.3$ keV event is clearly below this value (additionally, there are non-negligible astrophysical uncertainties in the value of the galactic escape velocity.) To further illustrate this point, we show in \Fig{fig:delta_neg225_fnfp_neg07} an analysis of a `Xe-phobic' dark matter candidate with $\delta = -225$ keV. It can be clearly seen in the halo-independent analysis (shown in the right panel for $m= 1.1$ GeV) that the third event of CDMS-II-Si can only be attributed to WIMPs traveling at speeds $v \simeq 1000$ km/s (in the lab frame), far above the galactic escape velocity. Notice that we have not included in our halo-independent analyses a term in the likelihood penalizing large unphysical halo speeds (as was done \eg in \cite{Feldstein:2014ufa}), which in this case would allow only two of the events observed by CDMS-II-Si to be attributed to dark matter. Also shown in \Fig{fig:delta_neg225_fnfp_neg07} is the SHM $\teta$ function with a normalization set to $\sigma_p = 2 \times 10^{-43}\rm{cm}^2$, a value which is allowed the $68\%$ CL region in the halo-dependent analysis. The halo-independent analysis clearly rejects this function at the $90\%$ CL, showing that, for this particular dark matter particle candidate, the SHM does not fit the CDMS-II-Si data well.

 \begin{figure*}
\center
\includegraphics[width=.49\textwidth]{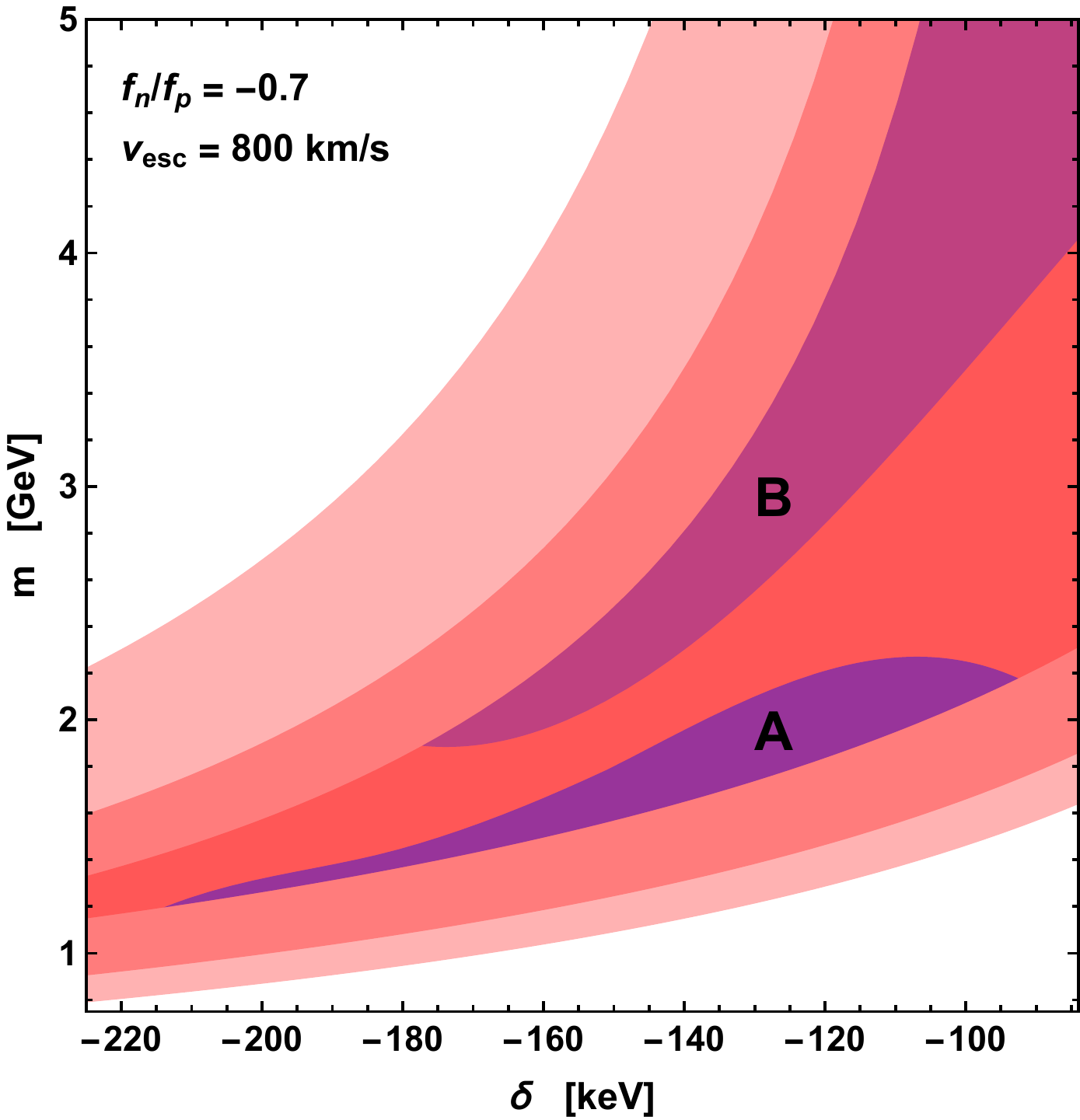}
\caption{\label{fig:exocomp}Values of $m$ and $\delta$ in the `Xe-phobic' model for which one (light pink), two (pink), and all three (redder pink) of the CDMS-II-Si events are below the galactic escape velocity. The two purple regions indicate the part of the redder pink region not excluded by current experiments. While still viable, the light purple, region `B', region provides a worse fit to the CDMS-II-Si data than the darker purple region, region `A' (see text for details).}
\end{figure*}

For strongly exothermic candidates, a small change in the particle mass leads to a considerable change in the range of recoil energies probed by acceptable values of $\vmin$. In \Fig{fig:delta_neg225_fnfp_neg07_omass}, we show the halo-independent analysis for the same interaction (\ie spin-independent with $f_n/f_p=-0.7$ and $\delta = -225$ keV) and a WIMP mass $m = 1.2$ GeV instead of $m = 1.1$ GeV. This small change in the mass eliminates the problem of requiring unacceptably large WIMP speeds. However, in this case the new PICO-60 90\% CL limit rejects the halo-independent CDMS-II-Si 90\% CL region, which would otherwise be allowed by all other bounds. Again, the SHM $\teta$ function with values of $\sigma_p$ allowed in the 90\% CL region in the halo-dependent analysis, lies outside the halo-independent 90\% CL confidence band. These examples clearly illustrate the point that one cannot continue to lower delta below $- 200$ keV with the hope of increasing the viability of a dark matter interpretation of the CDMS-II-Si events. 

For completeness, we show in \Fig{fig:exocomp} the viable parameter space in the $\delta-m$ plane for `Xe-phobic' models. The pink regions are where either one (light pink), two (pink), or all three (redder pink) events observed by CDMS can be induced by WIMPs traveling at speeds $v \leq 800$ km/s in the lab frame (a conservative choice for the galactic escape velocity~\cite{Green:2017odb}), namely where their recoils are kinematically allowed. The purple regions highlight the subset of the dark red region that cannot be ruled out by current direct detection experiments (\ie the viable parameter space where all three observed events can be due to WIMPs bound to the galactic halo). While the light purple region (region `B') is still viable, the minus log likelihood evaluated in the light purple region is significantly larger than that of the dark purple region (region `A'), indicating a worse fit to the data. This is because two of the observed events are relatively close in energy and thus, given that the halo function is monotonically decreasing, the data prefers models in which the $\vmin$ values associated with the two lowest observed recoils are lower than the $\vmin$ value of the highest energy event.

\Fig{fig:delta_neg200_fnfp1} and \Fig{fig:delta_neg225_fnfp_neg07} show that no viable parameter space for `Xe-phobic' interactions will remain if an experiment like LZ or PICO-250 does not find any dark matter signal (\ie the purple regions, both `A' and 'B', shown in \Fig{fig:exocomp} will be rejected). Notice that, even though the exposure of PICO-250 is much smaller than the exposure of LZ, PICO is highly sensitive to light exothermic WIMPs because fluorine is much lighter than xenon (in general exothermic scattering favors lighter target nuclei) and both PICO and LZ have comparable energy thresholds. In this regard, although argon is much lighter than xenon, the higher energy threshold of DarkSide-20k makes this experiment insensitive to light exothermic WIMPs.



\section{Conclusions\label{sec:conclu}}
We have presented here updated halo-dependent and halo-independent constraints on dark matter particle candidates that could explain the CDMS-II-Si data. We have studied candidates with isospin conserving and isospin-violating spin-independent interactions, with either elastic or exothermic scattering. We included constraints from PandaX-II, LUX (complete exposure), and PICO-60, as well as projected sensitivities for XENON1T, SuperCDMS SNOLAB Ge(HV), LZ, DARWIN, DarkSide-20k, and PICO-250.

The results presented show that both spin-independent isospin conserving and `Ge-phobic' ($f_n/f_p=-0.8$) interpretations of CDMS-II-Si are excluded at the $90\%$ CL. `Xe-phobic' ($f_n/f_p=-0.7$) interpretations, however, are still marginally viable after the recent PICO-60 result if the dark matter particle scatters exothermically with nuclei (with $\delta \lesssim -200$ keV), and can only be ruled out in the future by an experiment comparable to LZ or PICO-250. Although still marginally viable, the highly tuned nature of these models make a dark matter interpretation of the CDMS-II-Si very unlikely.

\bigskip

\textbf{Acknowledgments.} G.G. was supported in part by the US Department of Energy Grant DE-SC0009937. S.W. was supported by a UCLA Dissertation Year Fellowship. 

\bibliographystyle{JHEP}
\bibliography{lighthaloindep_update}
\end{document}